\documentclass[aps,twocolumn,amsfonts,amsmath,amssymb,superscriptaddress,showkeys]{revtex4-1}	

\usepackage[dvips]{graphicx}	%this shows the images -compiles slowly!

\usepackage{txfonts}
\usepackage{hyperref}   
\usepackage{breakurl}
\usepackage{microtype}        
\usepackage{color}
\usepackage{siunitx}
\usepackage{subfigure}

\graphicspath{{Figures/},.}

\newcommand{\ie}{i.e.,}
\newcommand{\eg}{e.g.,}

\newcommand{\proof}{\parindent=0pt \textbf{Proof.} }
\newcommand{\vect}[1]{#1}
\newtheorem{definition}{Definition}
\newtheorem{theorem}{Theorem}

\begin{document}

\title{Resilience of natural gas networks during conflicts, crises and disruptions}

\author{Rui~Carvalho}
\email{r.carvalho@qmul.ac.uk}
\affiliation{School of Mathematical Sciences, Queen Mary University of London, Mile End Road, London E1 4NS, U.K.}

\author{Lubos~Buzna}
\affiliation{University of Zilina, Univerzitna 8215/1, 01026 Zilina, Slovakia}

\author{Flavio~Bono}
\affiliation{European Laboratory for Structural Assessment, Institute for the Protection and Security of the Citizen (IPSC), Joint Research Centre, Via. E. Fermi, 2749 TP 480, Ispra 21027 (VA), Italy}
\affiliation{School of Mathematical Sciences, Queen Mary University of London, Mile End Road, London E1 4NS, U.K.}

\author{Marcelo~Masera} 
\affiliation{Energy Security Unit, Institute for Energy and Transport, Joint Research Centre, Westerduinweg 3,NL-1755 LE Petten, The Netherlands}

\author{David~K.~Arrowsmith}
\affiliation{School of Mathematical Sciences, Queen Mary University of London, Mile End Road, London E1 4NS, U.K.}

\author{Dirk~Helbing}
\affiliation{ETH Zurich, Clausiusstrasse 50, 8092 Zurich, Switzerland}
\affiliation{Risk Center, ETH Zurich, Swiss Federal Institute of Technology, Scheuchzerstrasse 7, 8092 Zurich, Switzerland}

%%%%%%%%%%%%%%%%%%%%%%%%%%%%%%%%%%%%%%%%%%%%%%%%%%%%%%%%%%%%%%%%
\begin{abstract}
Human conflict, geopolitical crises, terrorist attacks, and natural disasters can turn large parts of energy distribution networks offline. Europe's current gas supply network is largely dependent on deliveries from Russia and North Africa, creating vulnerabilities to social and political instabilities. During crises, less delivery may mean greater congestion, as the pipeline network is used in ways it has not been designed for.
Given the importance of the security of natural gas supply, we develop a model to handle network congestion on various geographical scales. We offer a resilient response strategy to energy shortages and quantify its effectiveness for a variety of relevant scenarios. In essence, Europe's gas supply can be made robust even to major supply disruptions, if a fair distribution strategy is applied.
\end{abstract}

\maketitle

Almost everything we do in the course of a day involves the use of energy. Yet, history has taught us that the threats to the security of supply come in unexpected ways~\cite{Yergin12,Levi13}.  Examples of unforeseen energy crises include the recent disputes between Russia and Ukraine over the price of natural gas ($2005$--$2006$, $2007$--$2008$, $2008$--$2009$)~\cite{Economist09}, the disruption of the oil and gas production industry in the US following Hurricanes Katrina and Rita ($2005$)~\cite{Mouawad05}, the terrorist attack on the Amenas gas plant that affected more than $10\%$ of Algerian production of natural gas ($2013$)~\cite{Chrisafis13}, and the supply shortage in March $2013$, when the UK had only $6$ hours worth of gas left in storage as a buffer~\cite{Plimmer13}. New vulnerabilities could come from cyber attacks to the infrastructure~\cite{Yergin12}, particularly in the case of state-driven attacks~\cite{Levi13}; be the result of prolonged uncertainty or inaction on energy security in the US or Europe~\cite{Schiermeier13}; or derive from an extended period of extremely volatile prices due to intense international conflict~\cite{Levi13}.  

Natural gas, a fossil fuel that accounts for $24\%$ of energy consumption in OECD-Europe~\cite{IEAGas12}, has been at the heart of these crises. Gas is expensive to transport, and this is done mainly over a pipeline network. The investments are large and are made with long-term horizons, often of decades, and the costs are covered by locking buyers into long-term contracts~\cite{Wright12}. Moreover, current infrastructure investments in Europe still derive from a historical  dependency on supply from Russia and North Africa~\cite{Moniz11}. This dependency leaves the European continent exposed to both a pipeline network that was not designed to transport large quantities of gas imported via Liquefied Natural Gas (LNG) terminals, and to the effects of political and social instabilities in countries that are heavily dependent either on the export of natural gas (\eg~Algeria, Libya, Qatar or Russia) or its transit (\eg~Ukraine). Hence, it is challenging to build infrastructure that will be resilient to  a wide range of possible crisis scenarios~\cite{HelbingNature13}.

In a crisis, less delivery may mean greater congestion. This is due to 
the breakdown of major transit routes or production losses in affected areas, which cause the supply network to be used in different ways from what it was designed for. Hence, the available resources cannot be distributed well with the remaining transport capacities~\cite{EnergyandEnvironment}~\cite{Voropai12,Lochner11}. This is why we need a method to handle congestion.

%%%%%% /FIGURE/ %%%%%%%
\begin{figure}
\centerline{\includegraphics[width=0.45\textwidth]{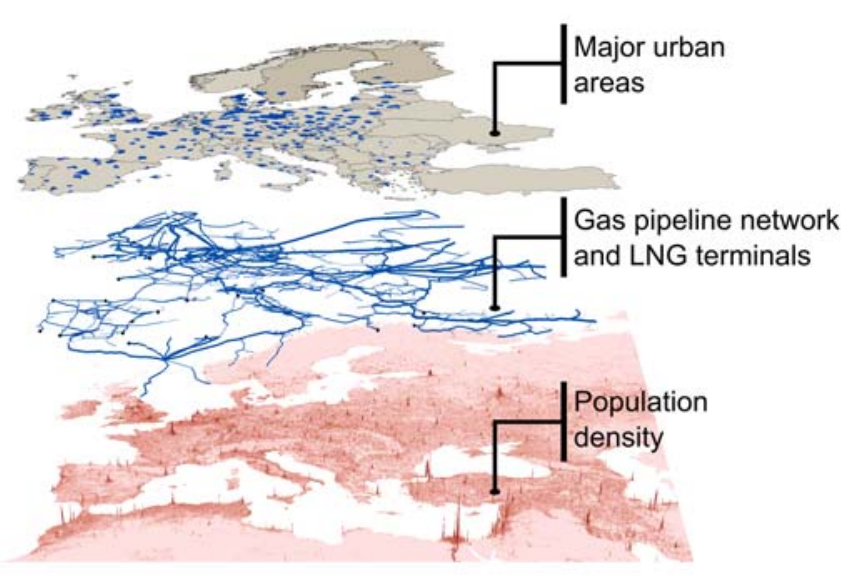}}
\caption{\label{fig:Figure_3d_maps} Spatial data layers involved in our analysis: population density (Landscan 2012); gas pipeline network and Liquefied Natural Gas (LNG) terminals (Platts 2011); and major urban areas (European Environment Agency and Natural Earth).}
\end{figure}
%%%%%% /FIGURE/ %%%%%%%

To manage the gas pipeline network during crises, we propose a decentralized model of congestion control that distributes the available network capacity to each route, without sacrificing network throughput~\cite{Kelly98,Bertsimas11,Carvalho12}. A central controller makes the system vulnerable both to attacks on the control centre and to delays and failures of the lines of communication through the network~\cite{Kelly98}. In contrast, a decentralized method is more resilient to failures because damage to the network has only a local effect and the need for communication is reduced.
To illustrate our model, we analyse the throughput of the present and planned pipeline networks across a range of different crisis scenarios at European, country and urban levels. The most challenging scenario corresponds to a hypothetical crisis with Russia with a complete cut-off of supply to Europe. We analyse how to alleviate the impact of such scenarios, by the identification of country groups with similar interests, which should cooperate closely to manage congestion on the network. This acknowledges that many of the $21$st century challenges, such as the management of energy grids and infrastructure networks~\cite{D'Souza10,Buldyrev10,Havlin12}, cannot be solved by technology alone, but do have a relevant behavioural or social component~\cite{Clark03,Carbone12,Levin13}.

\section*{Results}
\subsection* {Data set and model}
\begin{figure}
\begin{center}
\includegraphics[width=0.45\textwidth]{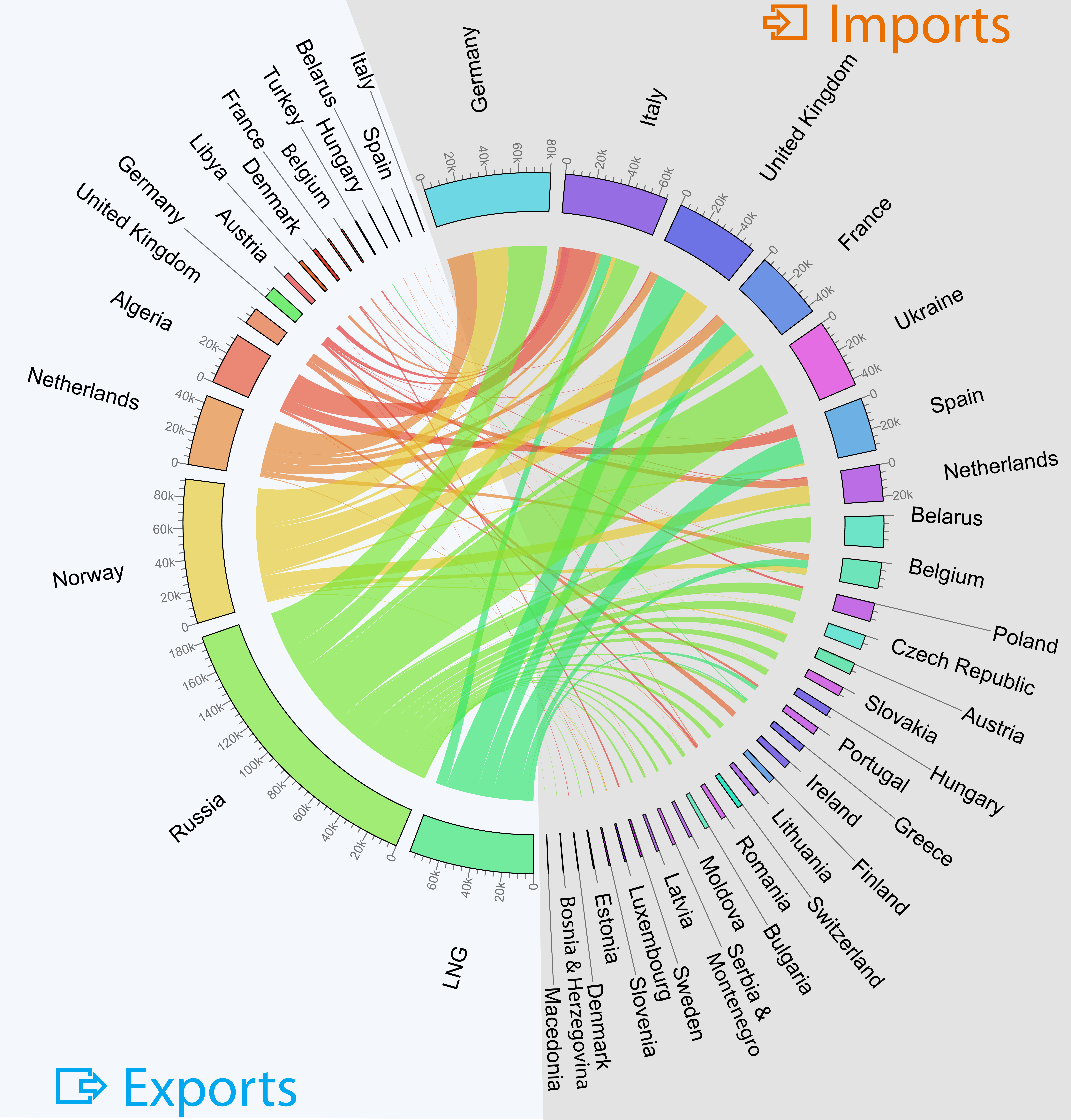}
\end{center}
\caption{\label{fig:flow_network} Natural gas imports by pipeline and via Liquefied Natural Gas (LNG) terminals in Europe during 2011 (million cubic meters). Gas exporting (importing) countries are on the left (right) of the image. For each exporting country, we show the breakdown of the volumes of gas exported annually, together with the importing countries served. For each importing country, we show the volumes of gas imported annually, together with the diversity of supply.}
\end{figure}

Our data set is organized in four layers (see Supplementary Information ``Databases''), three of which are shown in Figure~\ref{fig:Figure_3d_maps}. The first layer is the population density, which we compute from the $2012$ Landscan global population data set. The second layer is the European gas pipeline network and Liquefied Natural Gas (LNG) terminals, which we extract from the Platts $2011$ geospatial data set. This infrastructure is a spatial network, where nodes and links are geographically located, and links have capacity and length attributes. The third layer is defined by the urban areas in Europe with $100,000$ or more inhabitants, and we compile it from the \textit{European Environment Agency} and \textit{Natural Earth}. 
The fourth layer is the network of annual movements of gas via pipelines and of Liquefied Natural Gas (LNG) via shipping routes (see Figure~\ref{fig:flow_network}). We represent gas flowing from an exporting country $m$ (including LNG) to an importing country $n$, by a directed network with weighted adjacency matrix $T_{mn}$.  

%%%% FIGURE %%%%%%%
%%%%%% /FIGURE/ %%%%%%%

Gas enters the network at source nodes, is transported over long distances on the pipeline transmission network, and then passed to the distribution network that delivers it to consumers. Here we model only the transport of gas on the transmission network. To model consumption spatially, we first need a tessellation of each country into disjoint sets of urban and non-urban areas, such that the pipeline network in an area is associated with the population it serves. Urban areas are naturally defined by the boundary of their spatial polygons. 
We partition non-urban areas by a Voronoi tessellation with the gas pipeline nodes as generators, respecting country borders and excluding all urban areas~(see Supplementary Information ``The Model''). 

We assume that the flow of gas on each pipeline intersecting an urban polygon (\ie~the border of the urban area) is directed towards the centre of the urban area. For simplicity, we also assume that such pipelines supply the urban area from the closest node to the urban polygon that is located inside the urban area. Moreover, each non-urban area is defined by a Voronoi cell, and we assume that it is supplied by the cell generator node~(see Supplementary Information ``The Model''). 

To connect sink to source nodes with paths (see Contract Paths in Methods), we first go through each non-zero entry in the $T_{mn}$ transport matrix and link each sink node in an importing country $n$ to the $\Phi_{mn}=\min(10,s_m)$ closest nodes in an exporting country $m$, if $m$ is a country, or to all LNG terminals in country $n$, if $m$ is LNG, where $s_m$ is the number of gas pipeline nodes in an exporting country $m$.  

To allocate demand to individual paths, we start with the observation that the demand $T_{mn}$ of an importing country $n$ from an exporting country $m$ is proportional to the population of country $n$~\cite{Bettencourt07}. We next split the demand $T_{mn}$ among all source to sink paths between countries $m$ and $n$, proportionally to the population served by each sink node. We now have a value of demand associated with each path, and therefore with each sink node. Finally, we replace each path by a set of identical paths, each having the minimum demand on the network. This implies that all paths have the same demand, while doubling the demand on a path is equivalent to creating two identical paths with the original demand~(see Supplementary Information ``The Model'').

To begin integrating routing and congestion control, we first consider how to distribute the capacity $c_i$ of one single congested link over the $b_i=\sum_{j=1}^\rho B_{ij}$ paths that pass through the link, where $B$ is the link-path incidence matrix ($B_{ij}=1$ if link $i$ belongs to the path $r_j$ and $B_{ij}=0$ otherwise), and where $\rho$ is the number of paths on the network (see Table~1 of the Supplementary Information ``The Model'').
To find the exact routing for these paths, we apply an iterative algorithm that, for each source-sink pair, finds the path with minimum effective path length, where the effective link length is given by $\widetilde{l_i}=\left( \left\langle h_i\right\rangle /h_i\right)^{\alpha}l_i$, $l_i$ is the length of link $i$, $h_i=c_i/(1+b_i)$, and $\alpha=0.03$~(see Supplementary Information ``The Model''). 

We consider two baseline scenarios: the present and future networks. The present baseline scenario is the network that has been operational since $2011$; the future baseline scenario extends the present network by the planned and under construction pipelines. 
To determine the network effects of crises, we analyse a range of scenarios that consist in hypothetically removing exporting (\eg~Russia) or transit (\eg~Ukraine) countries from the baseline scenarios. The scenarios are, thus, identified by the baseline (present or future) and the hypothetically removed country. For example, the present Russia scenario is given by the present network excluding Russia, that is removing all entries in the transport matrix $T_{mn}$ that are movements of gas originating in Russia. Similarly, the future Ukraine scenario is determined by removing all Ukrainian nodes and links from the future network.

Broadly, there are three strategies to manage congestion~\cite{Frischmann12}. 
First, expanding the network capacity is the most obvious way to lower congestion. 
The EU has a plan to build major pipelines crossing the continent, that should lower European dependency on Russia (see planned pipelines in Figure~S1 of the Supplementary Information). Here, we include these planned pipelines in the future scenarios, but make no suggestions for extra infrastructure because the costs of expanding network capacity are high, and thus our focus is on how to best manage the existing and planned network capacity. Second, implementing congestion pricing is a way to cap the consumption of heavy users that cause network bottlenecks. Finally, by identifying groups of countries that have similar patterns of demand, we map a vast number of consumers to a relatively small number of communities that may be able to cooperate during crises~\cite{Ostrom03}.

We are aiming at controlling congestion in situations where the network has to perform a function for which it was not designed. For congestion control, we are using the \textit{proportional fairness} algorithm (see Methods), which is inspired by the way capacity is managed on the Internet~\cite{Kelly97,Kelly98,Srikant04}.
The main idea behind proportional fairness is to use pricing on the links in order to control congestion (see Methods and Supplementary Information ``Congestion Control''). Use of non-congested links is free up to a threshold, above which the cost that a path incurs for using a link increases linearly, but steeply, with the difference between link capacity and link utilization. Hence, paths that traverse many congested links pay a high cost for contributing to congestion, and thus get a smaller flow allocation than paths that avoid congestion. 
A flow is proportionally fair if, to increase a path flow by a percentage $\varepsilon$, we have to decrease a set of other path flows, such that the sum of the percentage decreases is larger or equal to $\varepsilon$.
We view the network as an optimizer and the proportional fairness policy as a distributed solution to a global optimization problem~\cite{Chiang07,Kelly11}. 

%%%% FIGURE %%%%%%%
\begin{figure}
\begin{center}
\includegraphics[width=0.45\textwidth]{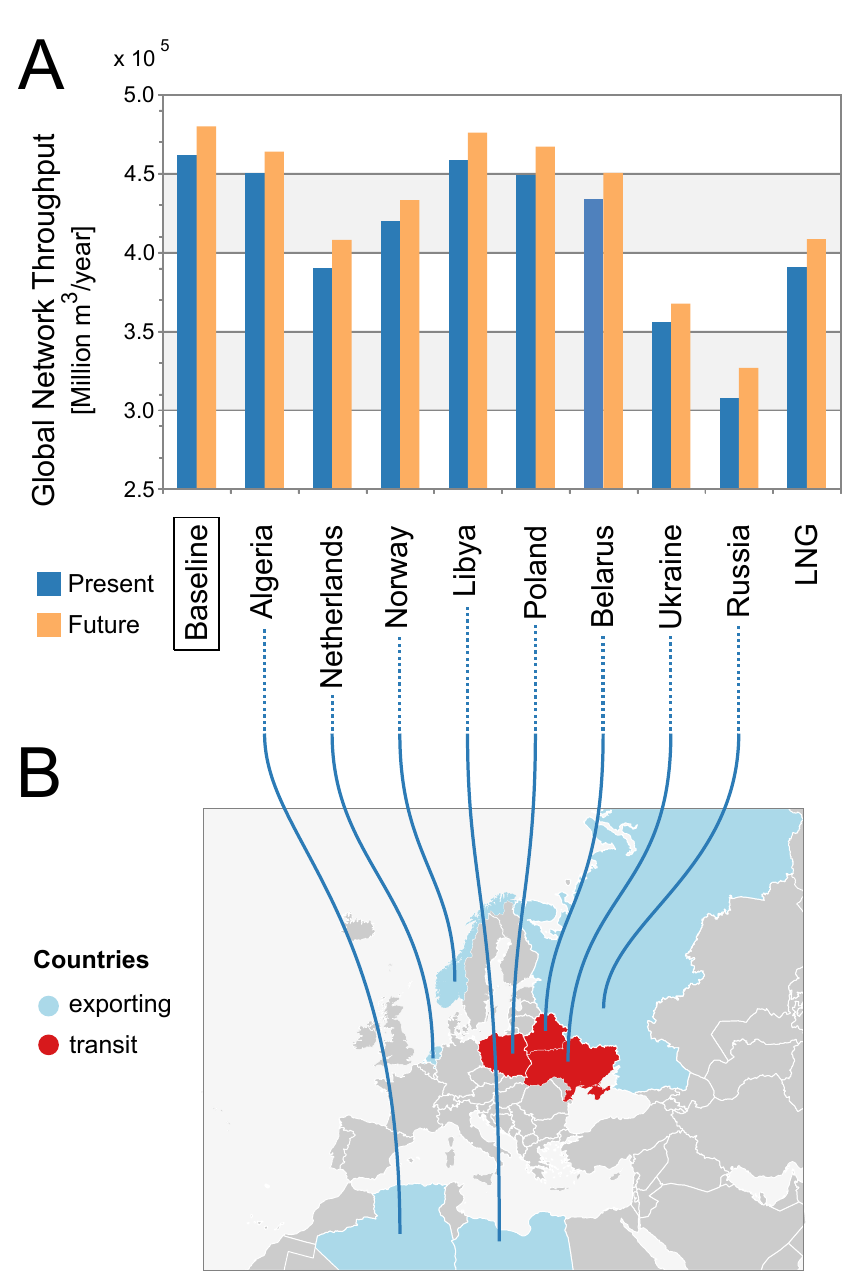}
\end{center}
\caption{\label{fig:network_throughput} Global network throughput by scenario. (A) A scenario is named after the country that is hypothetically removed from the network, and coloured in blue (orange) if the country is removed from the present (future) baseline scenario. (B) The country removed per scenario is coloured cyan (red) on the map, if it is an exporting (transit) country.
The total network throughput increases by $6.3\%$ from the present baseline to the future baseline scenario (\ie~when the future and planned pipelines are added to the present network).
The most challenging scenarios are the hypothetical removal of Russia, followed by Ukraine, the Netherlands and LNG. When Russia is removed from the network, the global network throughput falls by $32.7\%$ relative to the present baseline and by $28.1\%$ in relation to the future baseline.}
\end{figure}
%%%%%% /FIGURE/ %%%%%%%

%%%% FIGURE %%%%%%%
\begin{figure*}
\centering
\subfigure{\includegraphics[width=0.72\linewidth]{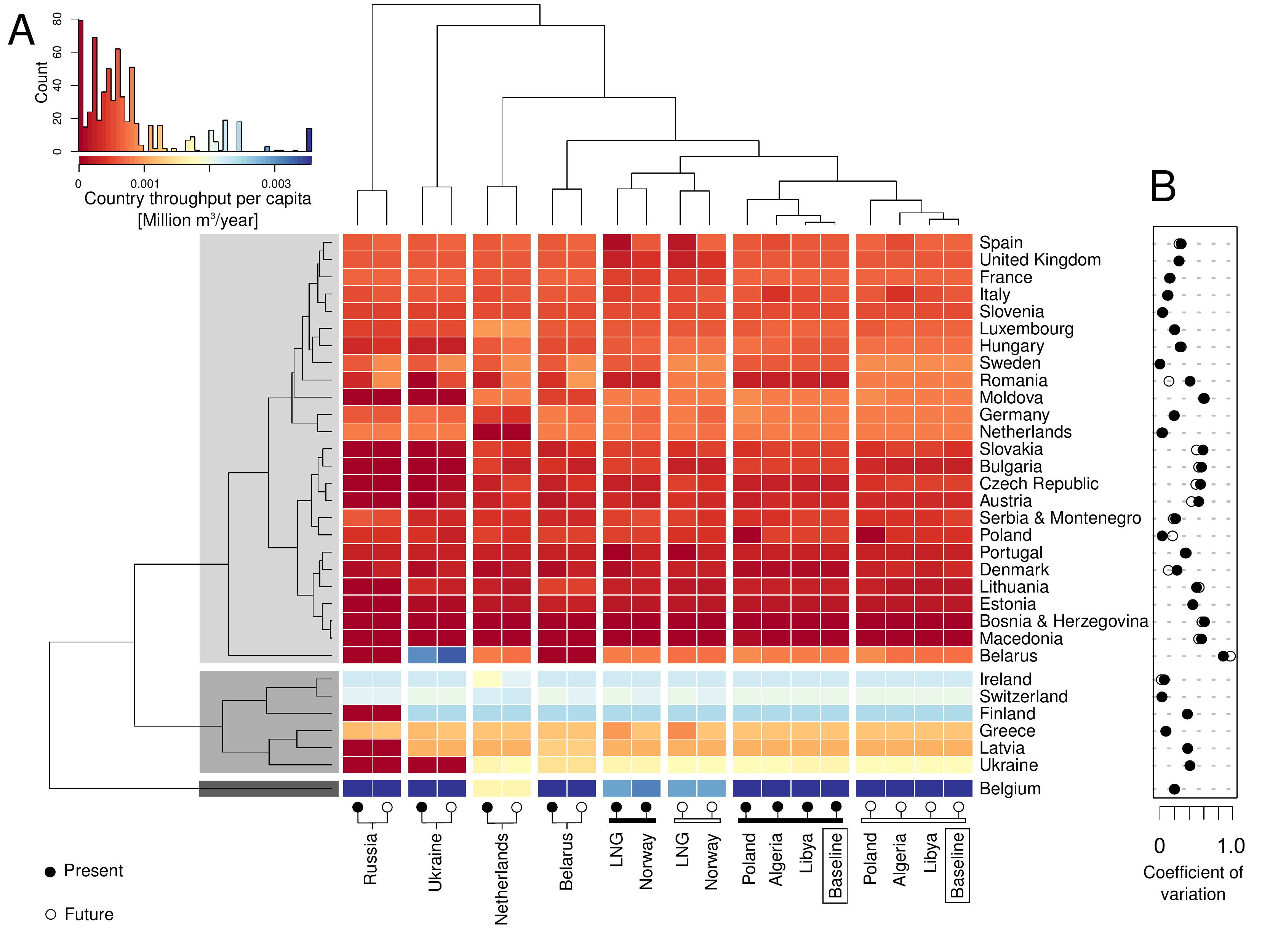} }
\subfigure{\includegraphics[width=0.72\linewidth]{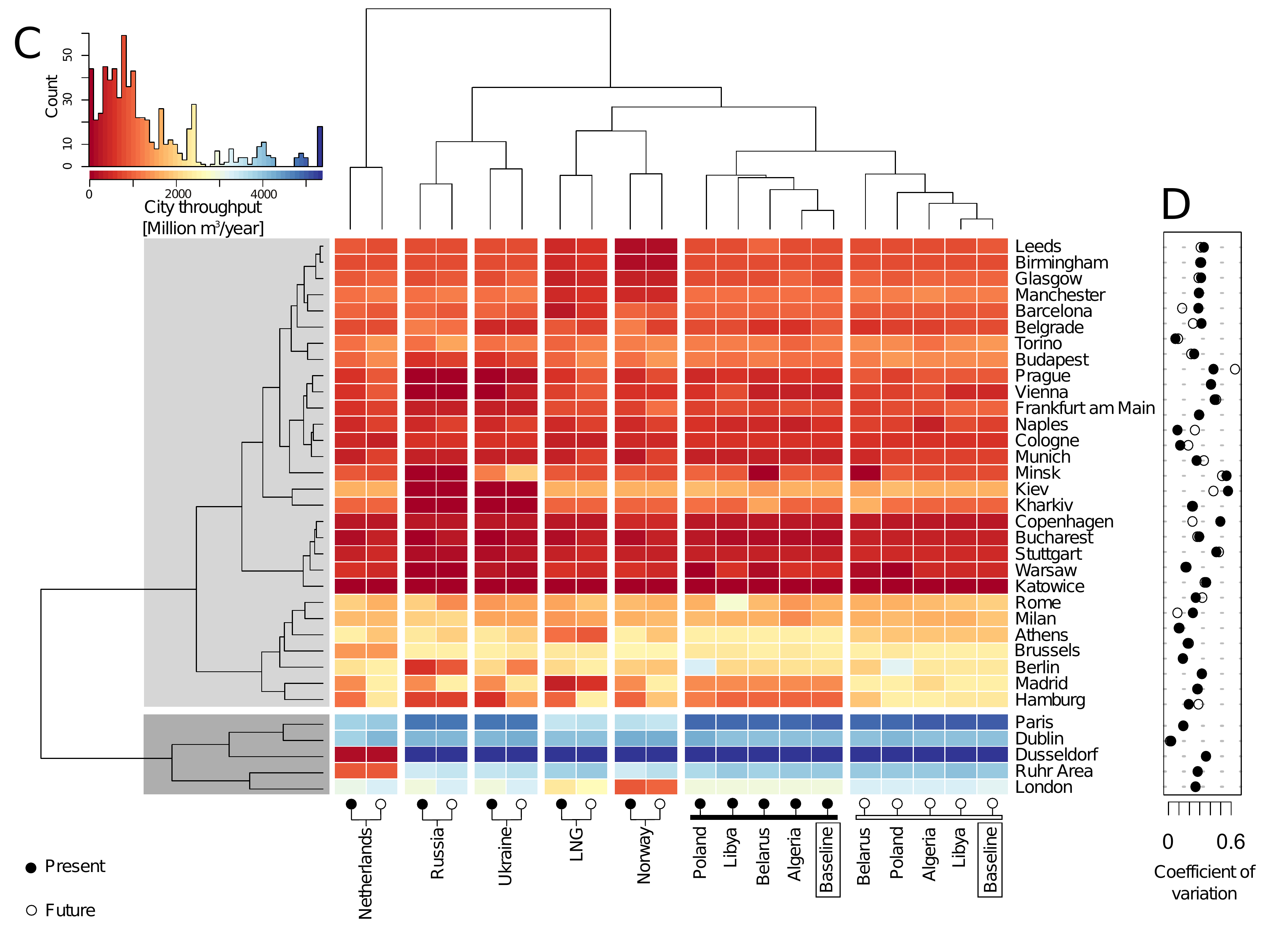} }
\caption{\label{fig:clustergram} Heat-map~\cite{Eisen98}, illustrating the variation of throughput across various scenarios and the effect of a scenario on the network. The dendrograms are computed using a hierarchical clustering algorithm with the Euclidean norm and average linkage clustering.
 (A) Heat-map of throughput at country level across various scenarios, allowing for a comparative analysis of the present versus future baseline scenarios, as well as of crises versus baseline scenarios;
(B) Coefficient of variation of throughput per capita of a country;
(C) Heat-map of throughput at urban level; (D) Coefficient of variation of throughput at urban scale. The gray areas denote groups of countries and urban areas that share common patterns of throughput across scenarios.}
\end{figure*}
%%%%%% /FIGURE/ %%%%%%%

\subsection*{Simulation Results}
For each scenario, we hypothetically remove the scenario country from the network and, if $m$ is an exporting country, remove row $m$ in the $T_{mn}$ transport matrix. Since the network topology and the flow network $T_{mn}$ depend on the scenario, we then re-compute the source-sink pairs, the demand of each pair, and we also replace every source-sink path with a number of identical paths, each having the minimum demand in the network. 
Finally, we apply the proportional fairness congestion control algorithm to the resulting network and paths. We assume that all countries are willing to cooperate, that is, adhere to the rules of the congestion control policy. To assess the effect of the range of scenarios, we then analyse the throughput at the scales of the European continent, countries, and of urban areas.

We compute the global network throughput, which is the sum of the throughput at all sinks (urban and non-urban), for all the scenarios. Our model reproduces successfully the expected consequences of removing the major source and transit countries from the network (see Figure~\ref{fig:network_throughput}).

We say that a country is resilient to crises if it combines high throughput per capita across scenarios with a low coefficient of variation of throughput. In addition, the network is considered resilient to a scenario if the vectors of country throughput per capita for the scenario and the baseline scenario are similar.
To start addressing the resilience of countries and the network to supply and transit crises, we study the signatures in the scenario space given by the country throughput per capita in each of the $20$ scenarios. Similarly, a scenario can be seen as a point in the $32$-dimensional space of country throughput. 
The heat-map in Figure~\ref{fig:clustergram}A shows the throughput per capita for each pair of countries and scenarios~\cite{Eisen98}. 

The country groups, determined by dendrograms and highlighted in gray, reflect a similar level of throughput per capita achieved across the scenarios. Countries belong to the high throughput per capita groups (highlighted in dark gray in the figure) due to a combination of effects:  diversity of supply; good access to network capacity (strategic geographical location); and a relatively small population (see discussion in Supplementary Information ``Results''). The coefficient of variation, shown in  Figure~\ref{fig:clustergram}B for present and future scenarios, measures the normalized dispersion of country throughput per capita using the mean as a measure of scale. Larger values indicate that the throughput accessible to a country varies across scenarios. Figure~\ref{fig:clustergram}B shows that countries in Eastern Europe have high coefficient of variation of throughput per capita in the scenarios where we hypothetically remove Russia or Ukraine. In other words, countries in Eastern Europe are still very much dependent on one single source country (Russia) and one major transit country (Ukraine). Unexpectedly, we observe a spillover effect from countries, such as Germany, which make large investments in infrastructure. These countries themselves seem to benefit less from such investments than some of their smaller neighbours.  The reason behind this spillover is that countries with plentiful access to network capacity provide routes for neighbouring countries to also access such capacity.

Figure~\ref{fig:clustergram}A can be read from left to right: the scenarios that cause the largest disruption appear on the left, and the most benign scenarios are on the right.  The present and future scenarios are clustered together when either Russia, Ukraine, the Netherlands, or Belarus are removed from the network, demonstrating that the new pipelines being built will only improve slightly the consequences of a hypothetical crisis with one of the major exporting countries (Russia or the Netherlands), or with a critical transit country (Ukraine or Belarus). It is thus very hard to change the consequences of such scenarios even by building new pipelines. 
 
We illustrate our model at a fine geographical scale in the heat-map of Figure~\ref{fig:clustergram}C, where we show the throughput for urban areas in Europe with $1.5$ million inhabitants or more, as the scenarios vary. 
The figure suggests possible classifications of cities into groups, highlighted in gray. 
We observe in Figure~\ref{fig:clustergram}D that the coefficient of variation is larger for cities in Eastern Europe than for cities elsewhere (except Berlin, Vienna and Dusseldorf). Note that Dublin is resilient to all scenarios because it is supplied from the UK, which we never removed from the network. Observe also that Austria gets most of its gas from Russia, and only a little from Norway, so Vienna is in a similar situation to Eastern European cities.

%%%% FIGURE %%%%%%%
\begin{figure}
\begin{center}
\includegraphics[width=0.45\textwidth]{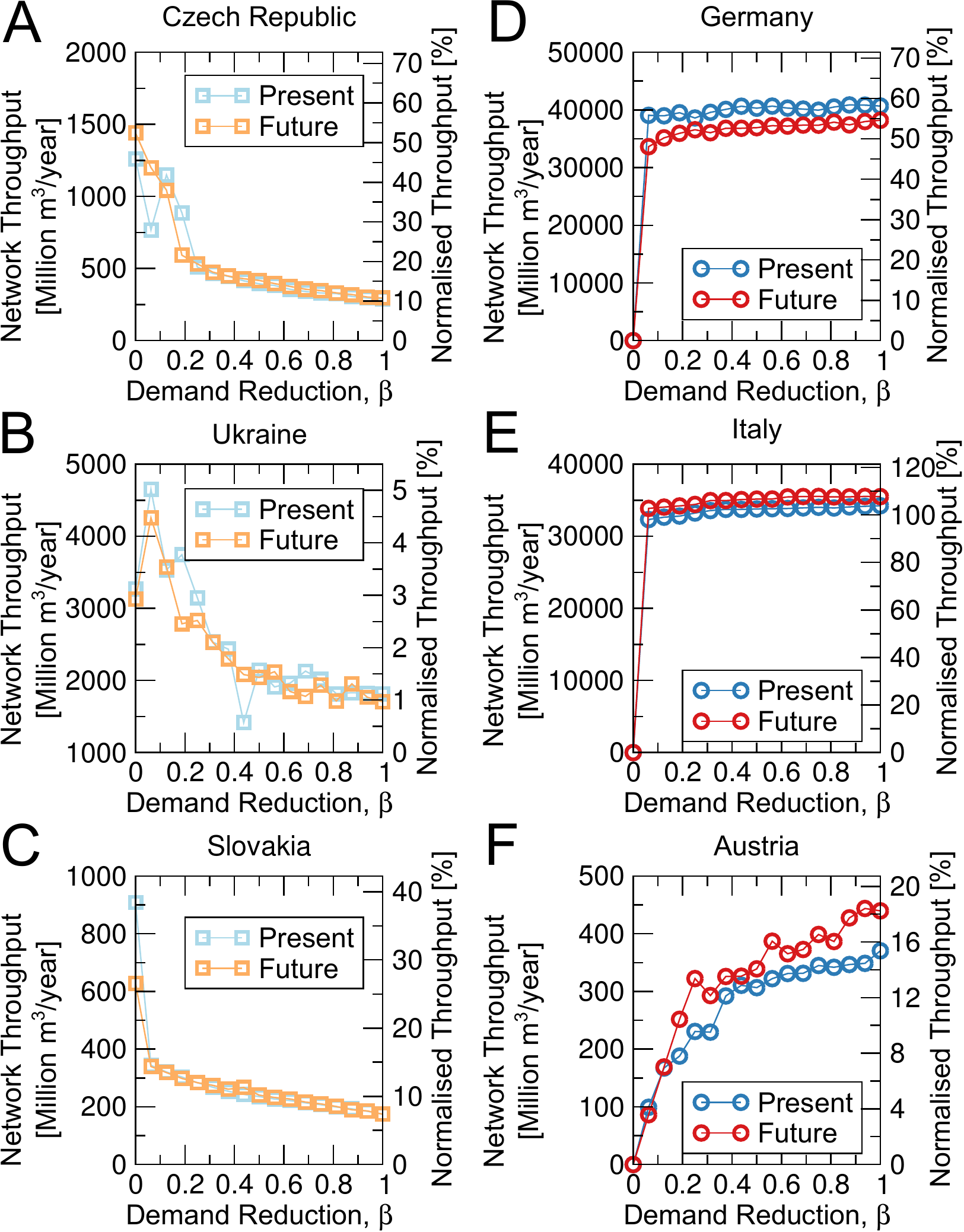}
\end{center}
\caption{\label{fig:west_vs_east} Network throughput of selected countries in a hypothetical crisis with Russia. The right axis shows the country throughput relative to the present baseline scenario.
To minimize the impact of the loss of Russian supply, we re-allocate paths that originate in Russia to Norway and the Netherlands (see Methods).
We then partition countries into two groups: group I  is composed of Eastern Europe (\url{http://eurovoc.europa.eu/100277}) together with Estonia, Finland, Greece, Latvia and Lithuania, and group II includes all other countries in our study. 
Group II countries have a demand of $\beta T_{mn}^\prime$, where $0\leq \beta \leq 1$.
Panels (A)--(C) show the throughput for selected group I countries (open squares), whereas panels (D)--(F) illustrate the throughput for group II countries (open circles). Panels (A)--(C) demonstrate that countries in group I benefit from curtailing the demand of countries in group II. In contrast, panels (D)--(E) show that some countries in group II are largely unaffected even when their own demand is curtailed considerably. Finally, panel (F) demonstrates that supply to Austria is dominated by the demand reduction prefactor, $\beta$. Indeed, Austria is crossed by routes from Norway and the Netherlands to group I countries, and these routes get a higher allocation of available capacity as Austrian demand decreases (\ie~as $\beta$ decreases).}
\end{figure}
%%%%%% /FIGURE/ %%%%%%%

Taken together, Figures~\ref{fig:clustergram}A--D illustrate the resilience of countries, urban areas and the network to the scenarios, by showing how countries and urban areas with similar reactions to different types of crises are grouped together by throughput or by its coefficient of variation, and how different scenarios are clustered by their effect on the countries and urban areas.

The most challenging scenario is a hypothetical crisis that would cut-off supply from Russia to Europe. To investigate how Europe could make use of its internal gas production to minimize the impact of such a crisis, we simulate and quantify the effect of replacing gas supply from Russia with supply from Norway and the Netherlands.
To do this, we start by creating two groups of countries. Group I is made of the countries that are heavily dependent on Russian gas, and is defined by Eastern Europe (\url{http://eurovoc.europa.eu/100277}) together with Estonia, Finland, Greece, Latvia and Lithuania. Group II is defined by all other countries in our study (see Supplementary Information ``Databases''). 
We consider a new scenario where Russia is removed from the network and the demand of countries in group I is rerouted to the Netherlands and Norway. To do this, we first create new paths linking each importing country in group I to Norway and the Netherlands (see Supplementary Information ``The Model'') and we update the matrix of gas flows to $T_{mn}^\prime$ (see Methods). Next, we apply a prefactor $0\leq\beta\leq 1$ to the values of the demand $T_{mn}^\prime$ of countries in group II. The effect of $\beta$ is to lower the utilization of the network by countries of group II that do not depend heavily on Russia. These countries typically have a high value of demand, and hence by curtailing their demand, there will be more capacity available to transport gas from Norway and the Netherlands to group I countries.
In Figure~\ref{fig:west_vs_east}, we observe that group I countries increase their access to network capacity as $\beta$ decreases. Group II countries, such as Austria, that are geographically on the main routes that link Norway and the Netherlands to group I countries, decrease their throughput as $\beta$ decreases. These countries are crucial: their throughput decreases as they share their network to benefit the more populous group I countries. In contrast, access to network capacity in routes supplying group II countries, such as Germany and Italy, is broadly unaffected, even as $\beta$ is lowered considerably, because routes from Norway and the Netherlands to group I countries use little network capacity from these group II countries. Despite the increase in throughput for countries in group I as $\beta$ decreases, Figure~\ref{fig:west_vs_east} shows the difficulty in replacing Russia by the Netherlands and Norway. Although we can hope to recover between $40$ and $50\%$ of the baseline throughput for the Czech Republic and Slovakia, we will only recover up to $5\%$ of the Russian supply to Ukraine and up to $20\%$ of the Austrian supply. 
 
\section*{Discussion}
Agreed political management processes are needed for crises scenarios, to guarantee supply to the most affected countries and urban areas and minimize the loss of gas by populations. Here, we propose a decentralized algorithm inspired by congestion control on the Internet, which would eliminate the need of improvisation and complicated, lengthy negotiations every time a crisis occurs. Such mechanism has a stabilizing effect because it lowers the resource deficiency of the most affected countries~\cite{Ostrom03,HelbingNature13}. We demonstrate how a wide range of scenarios impacts network throughput at global, country and urban levels, and how countries and urban areas react to scenarios of hypothetical crises. We show and quantify how countries that are heavily dependent on Russian supply can lower the impact of a crisis, if other countries accept to reduce their demand. Finally, our model tries to systematically compare alternative policy options during energy crises, using complex system models~\cite{Lempert02}.

In summary, Europe is not necessarily trapped and helpless during energy crises. The long-term interest in the sustainability of the gas industry makes governments and the industry likely to invest in rules and norms to enhance reciprocity and collective efforts during crises. Because the number of governments and companies ultimately involved in taking the decisions in Europe is relatively high, governments could implement decentralized solutions similar to the one we propose here, perhaps with a centralized control solution as backup. 
At its heart, energy security, like preparedness for future pandemics~\cite{Colizza07}, is about cooperation among nations~\cite{Yergin12}. To avoid European-wide crises, nations must cooperate to share access to their critical infrastructure networks.

\section*{Methods}
\label{sec:methods} 
Let $G=(V,E,c,l)$ be an undirected and connected weighted graph with no loops, node-set
$V$ and link-set $E=\{1,\ldots,\eta\}$. Each link $i$ has a capacity $c_i$ and a length $l_i$. The network has a set of $\rho$ paths connecting source to sink nodes.
All links of a path transport the same \textit{path flow}. Different paths can share a link, even to perform transport in different directions (\eg, during distinct time intervals). 

The relationship between links and paths can be described by the  \textit{link-path incidence matrix} $B$ as follows. Set $B_{ij}=1$ if the link $i$ belongs to the path $r_j$, and set $B_{ij}=0$ otherwise. Matrix $B$ 
has dimensions $\eta \times \rho$, and maps paths to the links contained in these paths. When $B$ is applied to a vector of path flows, the resulting vector with components $(Bf)_i=\sum_{j=1}^{\rho}B_{ij}f_j$ is the total flow on the links, or link throughput.
We say that a link is a \textit{bottleneck} if the sum of the path flows of paths that pass through it is equal to the link capacity. 
We assume that flows are elastic, that is that path flows are determined by the available network capacity.

\subsection*{Contract paths}
The pipeline contracts are for physical point-to-point transport on a given system over a \textit{contract path}~\cite{Strbac04}. The contract path is a route between a pair of source and sink nodes, such that gas flows from source to sink along that path and the transport costs are only incurred on links along that route.  

\subsection*{Proportional fairness congestion control: a primal algorithm}
A decentralized algorithm for congestion control (see Supplementary Information ``Congestion Control'') solves the system of coupled ODEs:
%%%% EQUATION %%%%%%%
\begin{equation}
\frac{d}{dt}f_j(t) =1-f_j(t) \sum_{i=1}^\eta B_{ij}\mu_i(t),
\label{eq:Methods_diff_eqs_primal}
\end{equation}
%%%% EQUATION %%%%%%% 
where the price on link $i$ is 
%%%% EQUATION %%%%%%%
\begin{equation}
\mu_i(t)=p_i\left (\sum_{j=1}^\rho B_{ij}f_j(t)\right ),
\label{eq:Methods_mu_total_price}
\end{equation}
%%%% EQUATION %%%%%%%
and the price function is given by
%%%% EQUATION %%%%%%%
\begin{equation}
p_i(y) = \frac{\max (0,y-c_i+\epsilon)}{\epsilon ^2},
\label{eq:Methods_price}
\end{equation}
%%%% EQUATION %%%%%%%
\subsection*{Proportional fairness congestion control: a dual algorithm}
Consider a system where the shadow prices vary gradually as a function of the path flows (see Supplementary Information ``Congestion Control''):
%%%% EQUATION %%%%%%%
\begin{equation}
\frac{d}{dt}\mu_i(t) =\sum_{j=1}^\rho B_{ij}f_j(t) - q_i(\mu_i(t)),
\label{eq:diff_eqs_dual}
\end{equation}
%%%% EQUATION %%%%%%% 
where
%%%% EQUATION %%%%%%%
\begin{equation}
f_j(t) = \frac{1}{\sum_{i=1}^\eta B_{ij}\mu_i(t)},
\label{eq:price_new}
\end{equation}
%%%% EQUATION %%%%%%%
and $q(\cdot )$ is the inverse of $p(\cdot )$. As $\epsilon\to 0$, the dual and primal algorithms become equivalent.

\subsection*{Rerouting the demand from Russia to the Netherlands and Norway}
When Russia is removed from the network, we reroute paths between group I countries and Russia to paths between group I countries and the Netherlands and Norway. To do this, we pair the new source and sink nodes as described in Supplementary Information ``The Model'', but we modify the $T_{mn}$ matrix of gas flows. The new $T_{mn}^\prime$ matrix is found by reallocating the demand from Russia for group I countries to the Netherlands and Norway, proportionally to the production of gas of these two exporting countries:
%%%% EQUATION %%%%%%%
\begin{equation}
\begin{aligned}
T_{(NO)n}^\prime&=a_{NO}T_{(RU)n}\\
T_{(NL)n}^\prime&=a_{NL}T_{(RU)n},
\label{eq:T_prime}
\end{aligned}
\end{equation}
%%%% EQUATION %%%%%%% 
\noindent where $a_{NO}=\frac{\sum_{j}T_{(NO)j}}{\sum_{j}T_{(NO)j}+T_{(NL)j}}$ and $a_{NL}=\frac{\sum_{j}T_{(NL)j}}{\sum_{j}T_{(NO)j}+T_{(NL)j}}$ are the normalised proportions of supply from Norway and the Netherlands, respectively.
\begin{acknowledgments}
We gratefully acknowledge the support of the RAVEN EPSRC project (EP/H04812X/1). L.B. was partially supported by VEGA (project 1/0339/13) and APVV (project APVV-0760-11). We thank ETHZ for granting access to the Brutus high-performance cluster.
\end{acknowledgments}

\clearpage 

\begin{center}
{\Large Supplementary Information: Resilience of natural gas networks during conflicts, crises and disruptions}
\end{center}

\makeatletter\renewcommand{\@biblabel}[1]{[S#1]}\makeatother
\renewcommand{\citenumfont}[1]{S#1}
\renewcommand{\thesection}{S\arabic{section}}
\renewcommand{\thetable}{S\arabic{table}}
\renewcommand{\thefigure}{S\arabic{figure}}
\renewcommand{\theequation}{S\arabic{equation}}
\setcounter{figure}{0}
\setcounter{table}{0}
\setcounter{equation}{0}
\setcounter{section}{0}

\section{Databases}

\begin{figure*}
\centerline{\includegraphics[width=1\textwidth]{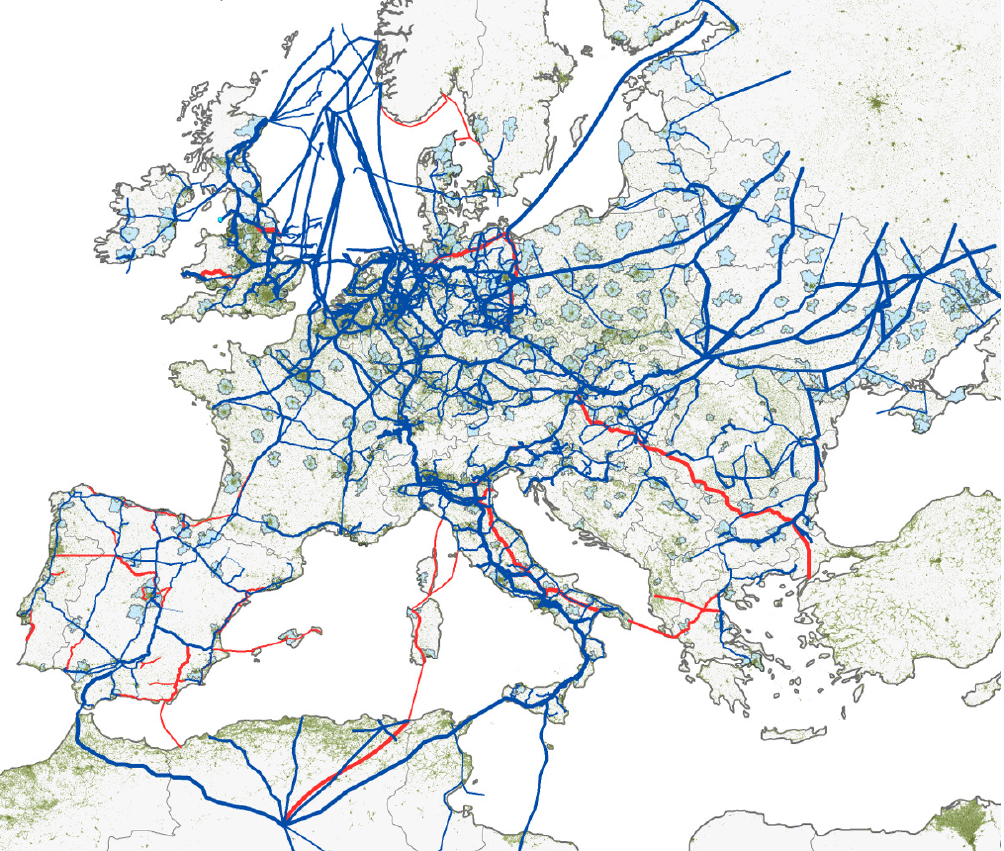}}
\caption{\label{fig:gas_pipeline_network} European gas pipeline network including part of North Africa. The present network is shown in dark blue, and the planned pipelines are shown in red. The population density is plot in dark green and Larger Urban Zones are indicated in cyan.}
\end{figure*}

\begin{figure*}
\centerline{\includegraphics[width=0.9\textwidth]{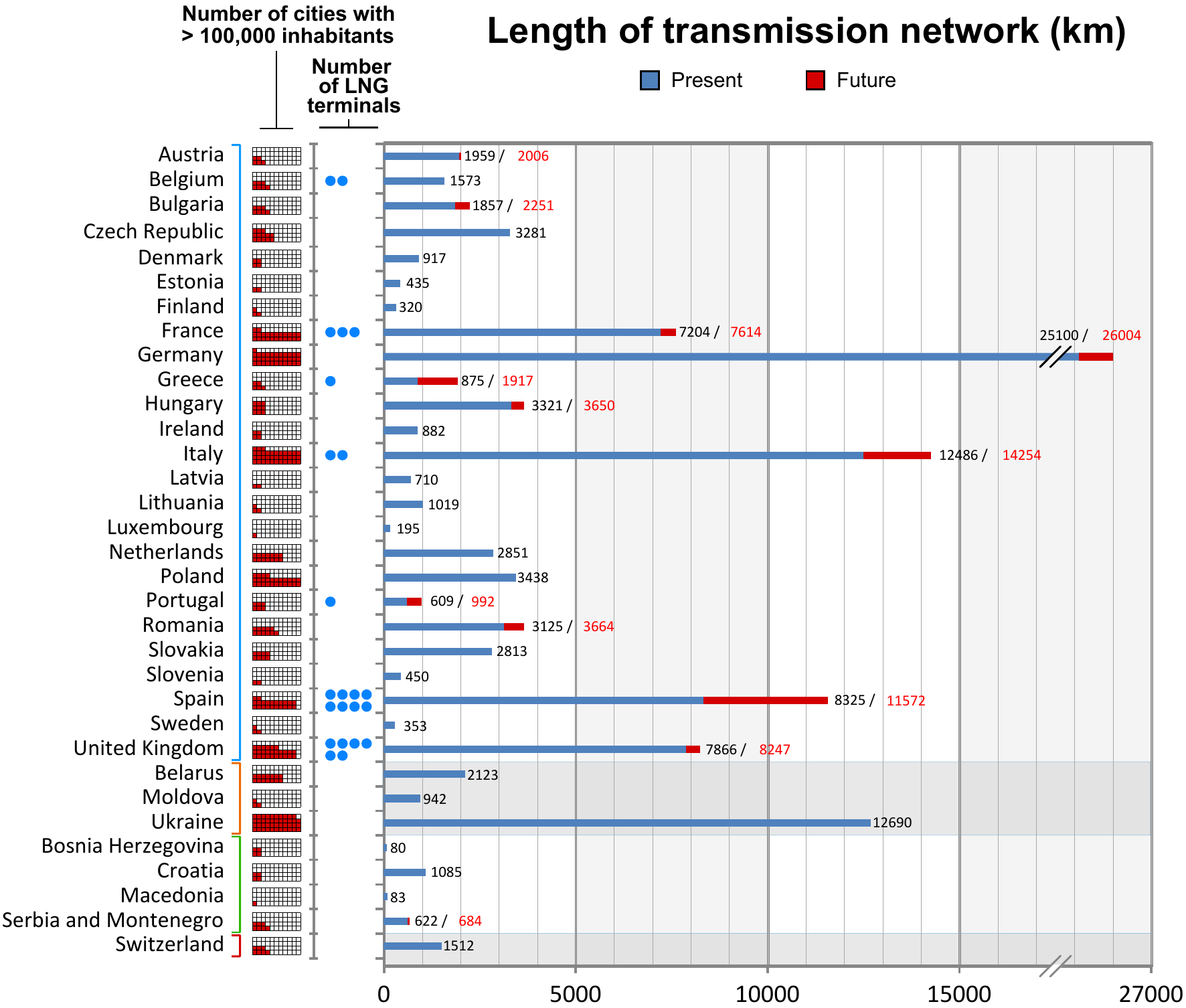}}
\caption{\label{fig:table1} Number of urban areas and Liquefied Natural Gas terminals, and length of the present and planned gas pipeline networks of the countries analysed.}
\end{figure*}

The data set is illustrated on Figures~\ref{fig:gas_pipeline_network} and~\ref{fig:table1} and is the result of compiling GIS and population databases into several layers.

\subsection{First layer: population}
We use the $2012$ LandScan~\cite{Slandscan02} high-resolution global population distribution data that estimates the population count with a spatial resolution of approximately $1$ km, or $30\times 30$ seconds of arc (see~\url{http://www.ornl.gov/sci/landscan/}).  

\subsection{Second layer: the gas pipeline network}
We compiled the European gas pipeline transmission network and the Liquefied Natural Gas (LNG) terminals from the $2011$ Platts Natural Gas geospatial data (see \url{http://www.platts.com/Products/gisdata}), including pipelines that are planned or under construction. The data set covers $25$ of the $27$ EU member states (except Malta and Cyprus), Belarus, Moldova, Western Russia, Ukraine (all part of the former USSR), Bosnia and Herzegovina, Croatia, Macedonia, Serbia (all part of the former Socialist Federal Republic of Yugoslavia), Algeria, 
Libya, Morocco, Tunisia (all part of the Maghreb), Norway, Switzerland and Western Turkey. 

%%%% FIGURE %%%%%%%
%%%%%% /FIGURE/ %%%%%%%

Similarly to electrical power grids, gas pipeline networks have two layers: transmission and distribution. The transmission network transports natural gas over long distances (typically across countries) and has a non-trivial topology. The distribution network is tree-like and comprises pipelines with smaller diameter that deliver gas to consumers. We extract the gas pipeline transmission network  as all the important pipelines with diameter $d \geq 15$ inches. To finalize the network, we add pipelines interconnecting major branches, so that the resulting network is connected. Network links are weighted by pipeline diameter and length. To simplify, we assume that gas can flow on both directions of a pipeline, although over different time periods. 
The compiled network has $2,649$ nodes (compressor stations, city gate stations, Liquefied  Natural Gas (LNG) terminals, etc.) connected by $3,673$ pipeline segments spanning $186,132$ km.

\subsection{Third layer: urban areas}
To avoid the controversy in the definition of an urban area~\cite[Ch IV, p 49]{SWorldBank05}, we considered only urban areas with $100,000$ or more inhabitants as defined by the Eurostat urban audit (see~\url{http://www.urbanaudit.org}). We are interested not just in the administrative boundaries of cities, but intend also to capture the surrounding areas that include a substantial share of the commuters into the city, since the gas pipeline infrastructure also supplies these peripheral urbanized districts. Note that the infrastructure network supplies directly the major urban areas, but may not intersect spatially with the built-up area of cities. 

Urban areas in the European Union member countries and candidate countries are defined by Eurostat as \textit{Larger Urban Zones} (\url{http://www.urbanaudit.org}), and the GIS files are provided by the European Environment Agency (\url{http://www.eea.europa.eu/data-and-maps/data/urban-atlas}). The city levels in non-EU countries are defined from remotely sensed data (see~\cite{SSchneider09} and~\url{http://www.naturalearthdata.com/downloads/10m-cultural-vectors/10m-urban-area/}). These city level areas are too small compared with the EU Larger Urban Zones. Hence, we define an urban area in non-EU countries to be the union of the third-level administrative divisions~(\url{http://www.gadm.org/}) that intersect the corresponding city level polygon. 
We have found $376$ urban areas with a total area of $723,957$ \si{\square\km}.

\subsection{Fourth layer: network of gas movements by pipeline and LNG}
The fourth layer is the network of annual movements of gas by pipeline and of Liquefied Natural Gas by ship into European terminals, collected from the International Energy Agency Natural Gas Information Statistics for $2011$~\cite{SIEA11} (see~\url{http://www.oecd-ilibrary.org/statistics}). This directed network is represented by the weighted adjacency matrix $T_{mn}$of gas transported from $m$ to $n$, where $m$ stands either for a gas exporting country or for Liquefied Natural Gas (LNG) terminals that supply an importing country $n$ (see Figure~$2$ of the main paper). We make use of
ISO alpha-$2$ country codes in $m$ and $n$ to denote individual countries (see \url{http://www.iso.org/iso/home/standards/country_codes.htm}), so that, for example, $T_{(RU)(FR)}$ is the amount of gas imported annually by France from Russia.

\section{The Model}
\subsection{Tessellation of urban and non-urban areas and location of source and sink nodes}
\begin{figure}
\centering
\includegraphics[width=0.9\linewidth]{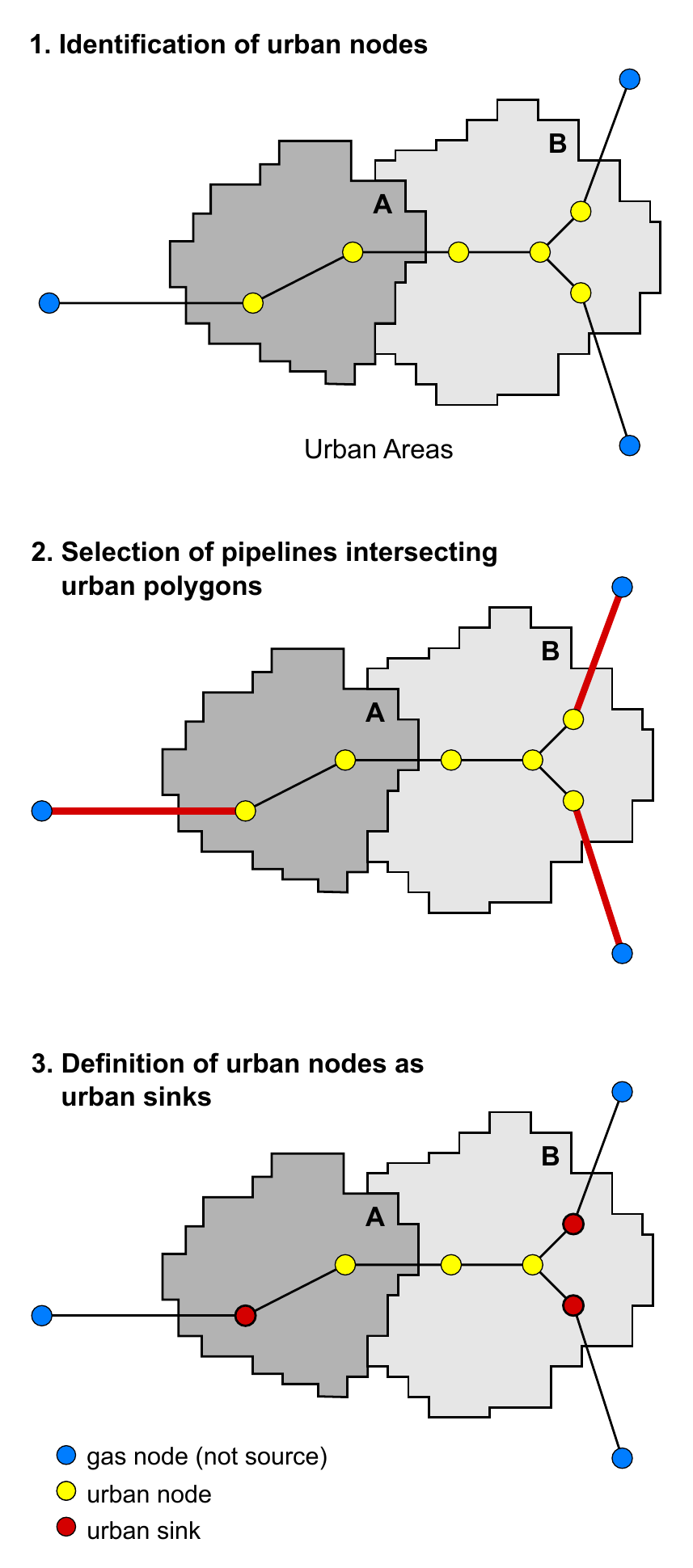}
\caption{\label{fig:UrbanSinks} Schematic figure illustrating the allocation of sink nodes. Urban sink nodes are shown in red, urban nodes that are not sinks are shown in yellow, and non-urban sink nodes are shown in blue.}
\end{figure}

\begin{figure*}
\centering
\includegraphics[width=0.9\linewidth]{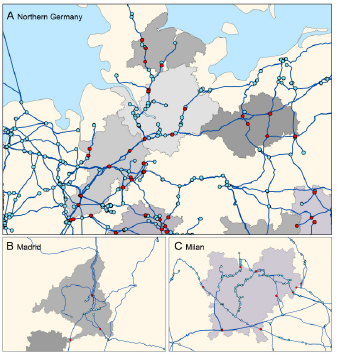}
\caption{\label{fig:city_sink_panels} Urban sink nodes are highlighted in red for the Larger Urban Zones of (A) Hamburg and North West Germany, (B) Madrid,  and (C) Milan.}
\end{figure*}

For simplicity, we consider that all nodes in a gas exporting country are source nodes. 
The partition of non-urban areas is such that all points within a given Voronoi cell are closer to their corresponding gas pipeline node than to any other node. 
If a gas pipeline node is inside an urban polygon, we call it an urban node, otherwise, we say the node is non-urban.
To simplify, we assume that an urban area is supplied by the pipeline links that cross its border and have a node inside its polygon. This node turns out to be also the urban node on the pipeline that is the closest to the border of the urban polygon, and is thus the first node that gas will cross along the pipeline when entering the urban area. Hence, we naturally say that the node is an urban sink, and we consider no other sink nodes along the pipeline for the given urban area. If an urban area polygon contains no gas nodes, we associate it to the closest gas node (urban or not) and say this node is a sink.
We exclude pipeline links that have both end nodes located inside urban areas. In other words, we only consider pipeline links that have one urban and one non-urban node (see Figures~\ref{fig:UrbanSinks} and~\ref{fig:city_sink_panels}). 

\subsection{How we pair sink and source nodes}
Source nodes are either located in an exporting country $m$, or at Liquefied Natural Gas (LNG) terminals. When $m$ stands for a country, we connect  by a path $r_{m,k,n,l}$ ($k=1,\dots,\Phi_{mn}$ and $l=1,\dots,t_n$) the $t_n$ sinks in an importing country $n$ to the $\Phi_{mn}$ closest source nodes in the exporting country $m$, where
%%%% EQUATION %%%%%%%
\begin{equation}
\Phi_{mn}=\left\{
\begin{array}
[c]{ll}%
\min(10,s_m) & \text{if }m\text{ is a gas exporting country}\\
g_n & \text{if }m\text{ is LNG}
\end{array}
\right. . \label{eq:phi}%
\end{equation}
%%%% EQUATION %%%%%%%
In other words, when $m$ is a country, we connect each sink node in an importing country $n$ to a maximum of ten source nodes in an exporting country $m$. When $m$ stands for LNG, we assume that sink nodes in an importing country $n$ are supplied from all the LNG terminals in country $n$ (see summary of the notation in Table~1).

%%%% FIGURE %%%%%%%
%%%%%% /FIGURE/ %%%%%%%

\subsection{How we define demand} 
We define the demand of a country to be the amount of gas imported over the gas pipeline network and Liquefied Natural Gas terminals, as given by the $T_{mn}$ matrix (see Figure~$2$ of the main paper), and the demand of a given geographical area to be the demand of the country weighted by the ratio between the area and the country populations. 
Since demand for energy is proportional to population~\cite{SBettencourt07}, we locate the sink nodes and associate them with the population they supply. 
When the area is urban, we split its total population equally among the sink nodes inside the urban polygon. If an urban area contains no gas nodes inside its polygon, we add its population to the population associated with the closest gas node. In non-urban areas, we associate the gas pipeline node at centre of a Voronoi cell with the population of the cell. Because each sink node in an importing country $n$ is connected by $\Phi_{mn}$ paths to source nodes in an exporting country $m$, each of these paths has a share of the demand $T_{mn}$ given by
%%%% EQUATION %%%%%%%
\begin{equation}
D_{mnl}=\frac{1}{\Phi_{mn}}\frac{Z_{nl}T_{mn}}{z_n}
\label{eq:demand}
\end{equation}
%%%% EQUATION %%%%%%%c
where $Z_{nl}$ is the population associated with sink node $l$ of importing country $n$, $z_n$ is the population of importing country $n$, $T_{mn}$ is the volume of gas imported by an importing country $n$ from an exporting country $m$, and the number $\Phi_{mn}$ of paths from an exporting country $m$ to each sink node is given by equation~(\ref{eq:phi}) (see summary of the notation in Table~1). 

We next express the demand of a path in units of the minimum demand on the network. To do this, we note that $D_{mnl}$ is independent of $k$, and  we replace  path $r_{m,k,n,l}$ having demand  $D_{mnl}$ by $\overline{D}_{mnl}$ paths identical to $r_{m,k,n,l}$, each with demand $\min(D_{mnl})$, where
%%%% EQUATION %%%%%%%
\begin{equation}
\overline{D}_{mnl}=\left \lfloor \frac{D_{mnl}}{\min(D_{mnl})}\right \rfloor,
\label{eq:normalized_demand}
\end{equation}
%%%% EQUATION %%%%%%%
where $\left \lfloor \cdot \right \rfloor$ is the largest integer not greater than $\cdot$. We note that the demand of a sink node from a source node is now proportional to the number of paths connecting the source and sink pair. 

The path notation $r_{m,k,n,l}$ has been useful so far to locate the origin and destination of the paths, but the congestion control algorithm uses matrix multiplication, and it is simpler from now on to index paths in the network by an integer. To do this, we loop through all pairs of exporting and importing countries with a non-zero entry in the $T$ matrix and re-label each of the $\overline{D}_{mnl}$ source-sink paths identical to $r_{m,k,n,l}$ by the new index.
In other words, for each pair of importing-exporting countries, we go through the $\overline{D}_{mnl}\Phi_{mn}  t_n$ shortest paths that connect source to sink nodes (see Table~1), and we index all paths in increasing order of first $m$, then $n$ and finally $k$. 
Now that we have allocated the source to sink paths, we update the number of paths on the network $\rho=\sum_{i=1}^\nu \sum_{j=1}^\nu \sum_{l=1}^{t_n} \overline{D}_{mnl}\Phi_{mn}\widehat{T}_{mn}$, where $\widehat{T}_{mn}=1$ if $T_{mn}$ is positive and zero otherwise, and we write $r_j$ to denote path $j$, where $j=1,\dots,\rho$.

\subsection{The problem with shortest path routing}
\begin{figure*}
\centerline{\includegraphics[width=1\textwidth]{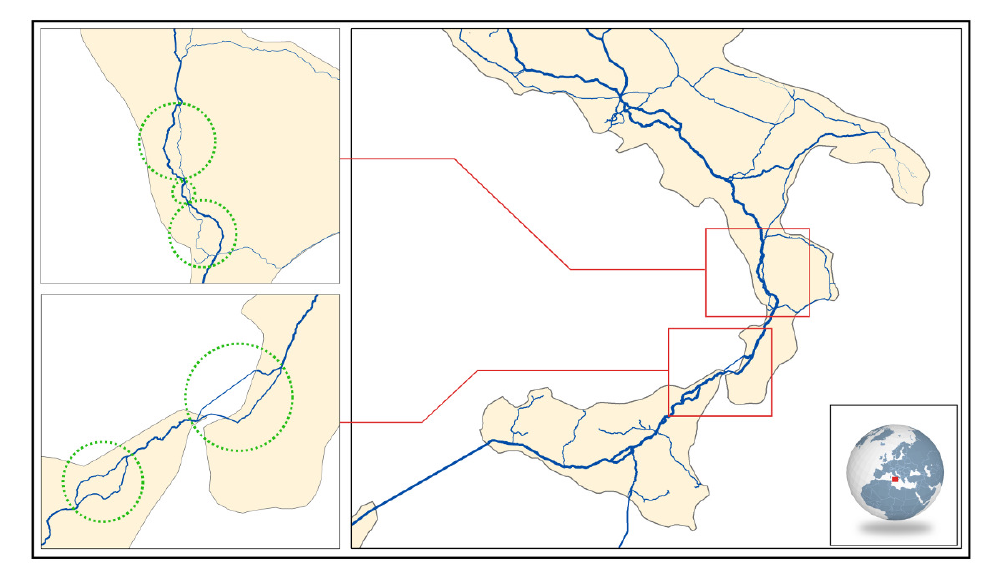}}
\caption{\label{fig:foot_of_Italy_parallel_pipes} Detail of the gas pipeline network in Italy, showing the presence of parallel routes.}
\end{figure*}

The pattern of route intersection determines how much the paths condition each other in their sharing of network links, and the capacity of links limits how much can be transported locally. If the network is not congested, transport over the geographical shortest paths minimizes the costs. In contrast, shortest path routing in congested networks can be inefficient, because it may cause congestion at a few overloaded links, while avoiding alternative routes that are only slightly longer but have higher capacity.
Moreover, routing over shortest paths in gas pipeline networks makes the effect of congestion even worst. Indeed, parallel routes with similar capacity are often available, but only one of these routes is the shortest path (see Figure~\ref{fig:foot_of_Italy_parallel_pipes}), and hence the network capacity is largely underused.

%%%% FIGURE %%%%%%%
%%%% FIGURE %%%%%%%

\subsection{How we determine the source to sink paths}
\begin{figure}
\begin{center}
\includegraphics[width=0.48\textwidth]{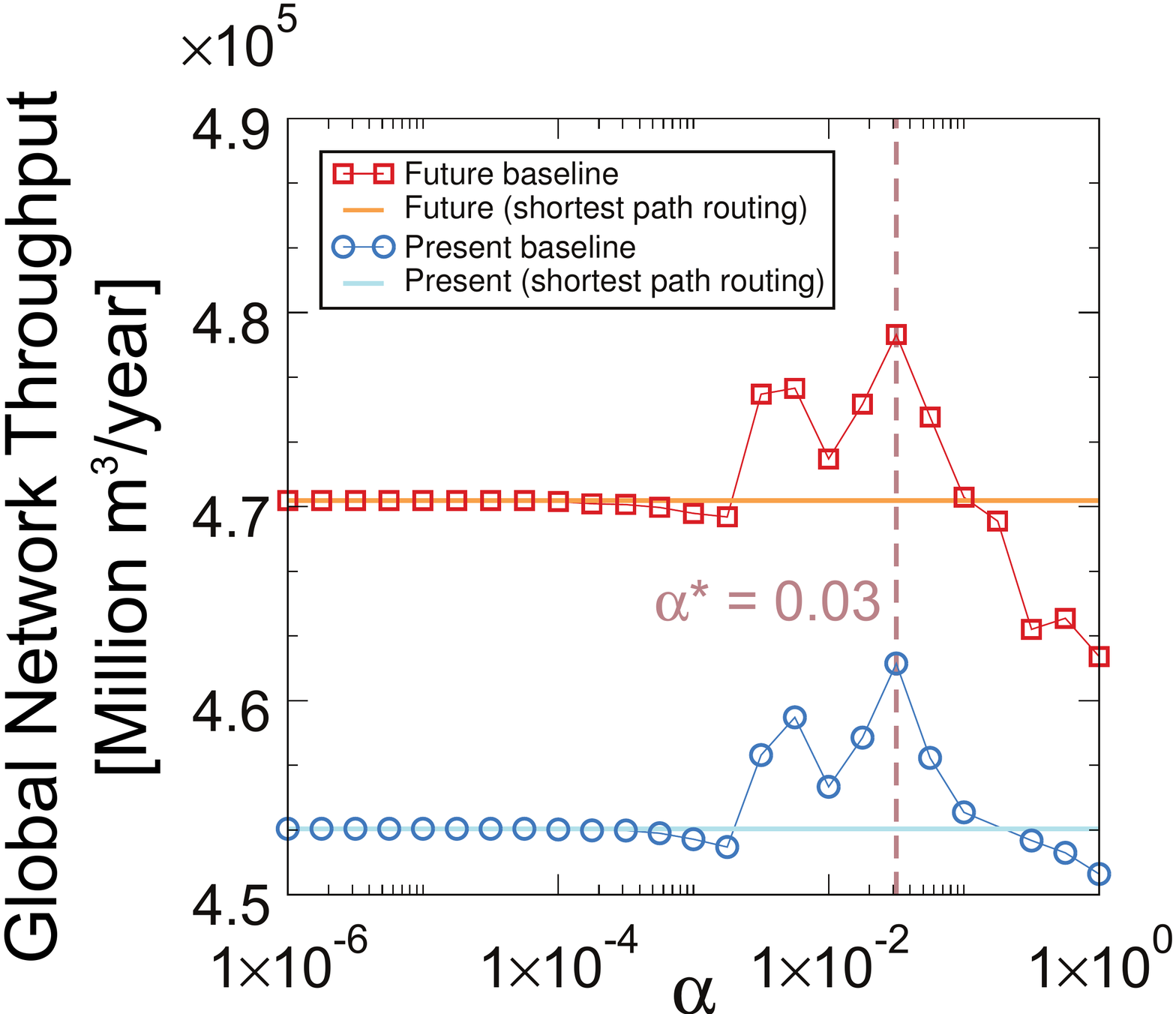}
\end{center}
\caption{\label{fig:ALPHA_parameter} Plot of the global network throughput in the present and future baseline scenarios when we apply the heuristic routing algorithm of Equation~(\ref{eq:link_weights}). Horizontal lines are a guide for the eye and show network throughput in the present and future baseline scenarios with shortest path routing. We choose the value $\alpha=0.03$ that maximizes the global throughput.}
\end{figure}

To begin integrating routing and congestion control, we consider first how to distribute the capacity $c_i$ of a congested link $i$ over the $1+b_i=1+\sum_{j=1}^\rho B_{ij}$ paths that pass through the link when we add a new path through $i$, where $B$ is the link-path incidence matrix ($B_{ij}=1$ if the link $i$ belongs to the path $r_j$ and $B_{ij}=0$ otherwise).
An equitable way to divide the capacity on the link is to assign a path flow of $h_i=c_i/(1+b_i)$ to each of the $1+b_i$ paths. Intuitively, $h_i$ is the slice of capacity allocated in a fair way to each of the $b_i$ paths that share the capacity $c_i$, and the split is viewed as a fair outcome~\cite{SFehr03}. Moreover, $1/h_i$ can be interpreted as a simple measure of network congestion, since it has a maximum at the most congested link~\cite{SCarvalho12}. Hence, we combine routing and congestion through an effective link length:
%%%% EQUATION %%%%%%%
\begin{equation}
\widetilde{l_i}=\left(  \frac{\left\langle h_i\right\rangle }{h_i}\right)^{\alpha}l_i,
\label{eq:link_weights}
\end{equation}
%%%% EQUATION %%%%%%%
where $\left\langle h_i\right\rangle$ is the average of $h_i$ over all network links, $l_i$ is the length of link $i$, and $0\leq \alpha<1$. Whereas we weight links by their length $l_i$ in the calculation of geographical shortest paths, we now weight each link by $\widetilde{l_i}$ in the calculation of weighted shortest paths. Thus, a link becomes less attractive (its effective length is increased) if it is more congested than the average. We find that the global network throughput is maximized for 
$\alpha=0.03$ (see Figure~\ref{fig:ALPHA_parameter}), and thus we use this value in the simulations.

We define the effective path length $\overleftrightarrow{l_j}$ of path $j$ as the sum of the effective lengths of each of its links. We interpret the sum of link weights on a path as a penalty, which we then use to reroute paths iteratively via the following heuristic~\cite{SSchneider11}. We  i) go through each source and sink node pair and find a new path $j$ connecting the two nodes; ii) if this new path has lower value of $\overleftrightarrow{l_j}$ than the previously found path, then it replaces the existing source to sink path; iii) we recompute the weights $\widetilde{l_i}$ for all links on the new paths and repeat the procedure for all paths, until it has been executed $20$ times (we found that the solution is does not change significantly when the number of iterations is larger than $20$).

\section{Congestion Control} 
How should we allocate scarce network resources to competing paths so as to manage network congestion? There are two mechanisms at play in such allocation. On one hand, maximizing the flow transported on the network may lead to some paths being assigned a zero share of network capacity, and hence zero path flow. These paths are effectively blocked from using the network, and hence the flow allocation is unfair. On the other hand, allocations that share network capacity fairly are known to deliver low throughput and are thus inefficient~\cite{SCarvalho12}. Hence, a good solution to the problem of congestion control aims at a trade-off between efficiency and fairness.

How should we generalize equation~(\ref{eq:link_weights}) when paths pass through several congested links on the network? Our intuitive notion of fairness breaks down on networks, because paths typically cross several congested links and hence share the capacity of these links with other paths.
Roughly, a solution is to allocate path flows iteratively, such that at each iteration we increase all path flows that do not pass through existing bottleneck links by the slice of capacity that is found by sharing equitably the capacity of the most congested links. The slice of capacity available to each path is the smallest on the most congested links, that is, the ratio $h$ is the smallest on these links. 
A procedure to do this finds one link $i$, with the smallest ratio $h_i$ and increases all path flows by $h_i$. Such procedure distributes parsimoniously the capacity of link $i$ among the paths that pass through the link, and increases all unsaturated path flows by $h_i$. The procedure then fixes the path flows of the paths that cross links with capacity $c_i$, and decreases the capacity of links crossed by these paths by the amount of flow fixed. This creates a residual network, on which the procedure is then repeated, such that path flows are saturated and the capacity available at the links they cross is updated at each iteration. The procedure is repeated until all paths in the network are saturated. Such allocation is known as \textit{max-min fair}\cite{SBertsekas92,SCarvalho12}, a name that comes from the way that path flows with minimum allocation are maximized by splitting equitably the capacity at the bottleneck links in an iterative process. The max-min fair allocation is such that to increase a path flow we have to decrease another path flow that is already smaller.

%%%% FIGURE %%%%%%%
%%%% FIGURE %%%%%%%

The major limitation of the max-min fair method is that network throughput is low compared to max-flow. To understand the mechanism behind this, we have to look at how both max-min fair and max-flow allocate path flows. The efficient allocation (max-flow) privileges short, over long paths that pass through several bottleneck links. Long paths take up capacity from other paths at each bottleneck, but only contribute to network throughput at the sink node. Hence, network throughput is maximized by minimizing the share of capacity to the long paths that pass through many bottlenecks, so that shorter paths can get a higher allocation of capacity and thus provide a higher contribution to network throughput. On the other hand, the max-min fair allocation shares the capacity of bottlenecks among the paths that pass through them. Thus, unlike max-flow, max-min fair allocations do not restrict the amount of network capacity that long paths can consume and are often inefficient. This limitation prompted the search for a trade-off between max-min fairness and max-flow, which would still distribute network capacity in an equitable way, and thus \textit{proportional fairness} appeared in the late $1990$s.

\subsection{Proportional Fairness}
Both proportional fairness and max-min fairness share the capacity $c_i$ of a single link among $N$ paths in a fair way, so that each path gets a path flow of $c_i/N$, but the two allocations are distinct when operating on a network. 

\begin{definition}
A vector of path flows ${\vect f^\ast}=(f_1^\ast,\ldots,f_\rho^\ast)$ is \textit{proportionally fair} if it is feasible and if for any other feasible vector of path flows ${\vect f}$, the sum of proportional changes in the path flows is non-positive~\cite{SKelly98,STan99}:
%%%% EQUATION %%%%%%%
\begin{equation}
\sum_{j=1}^{\rho}\frac{f_j-f_j^\ast}{f_j^\ast}\leqslant 0.
\label{eq:proportional_fairness}
\end{equation}
%%%% EQUATION %%%%%%%
\end{definition}
If there were no capacity constraints, equation~(\ref{eq:proportional_fairness}) would be verified when $f_j^\ast=\infty$ for all $j=1,\ldots,\rho$. The capacity constraints imply that a flow allocation $f^\ast$  is  proportionally fair if all other feasible vector of path flows $f_j = (1+\delta_j)f^\ast_j$, for $\delta\in\mathbb{R}^\rho$ where $j=1,\ldots,\rho$, verify that the aggregate of percent changes $\sum_{j=1}^\rho\delta_j$ is non-positive.
 
\begin{theorem} 
The unique set of feasible paths flows that maximizes the function $U(\vect f )=\sum_{j=1}^\rho \log(f_j)$ is proportionally fair.
\end{theorem}
\proof The proof given here is a direct application of the properties of convex functions~\cite{SGuler10, SSrikant04} and global maxima of a function (a sketch of the proof is given in~\cite{SKelly97}). First, observe that the set of feasible path flows is compact (closed and bounded) and convex. The functions $\log(f_j)$ are strictly concave, and thus $U(\vect f )$ is strictly concave, since it is the sum of strictly concave functions. Thus $U(\vect f )$ has a unique global maximum. Second, note that the tangent plane at any point of a convex (concave) function lies below (above) the graph of the function. Hence, since $U(\vect f )$ is concave:
%%%% EQUATION %%%%%%%
\begin{equation}
\nabla U(\vect f)\cdot (\vect g-\vect f)\geqslant U(\vect g)-U(\vect f).
\label{eq:concave_function}
\end{equation}
%%%% EQUATION %%%%%%%
Now assume that $\vect f$ is a proportionally fair allocation. Then, $\nabla U(\vect f)\cdot (\vect g-\vect f)\leqslant 0$ from equation~(\ref{eq:proportional_fairness}), and thus $U(\vect f)-U(\vect g)\geqslant 0$ from equation~(\ref{eq:concave_function}) for  all other feasible $\vect g$. Hence, $\vect f$ is a global maximum of $U$. Conversely, assume that the function $U$ has one global maximum at $U(\vect f)$. Then, 
%%%% EQUATION %%%%%%%
\begin{equation}
\nabla U(\vect f)\cdot (\vect g-\vect f)=\lim_{t\rightarrow 0^+}\frac{U(\vect f+t(\vect g-\vect f))-U(\vect f)}{t}\leq 0,\nonumber
\end{equation}
%%%% EQUATION %%%%%%%
and thus the flow allocation is proportionally fair.
$\Box$

\begin{theorem} If a vector ${\vect f^\ast}=(f_1,\ldots,f_\rho)$ of path flows is proportionally fair, then each path will pass through a bottleneck. 
\end{theorem}
\proof To see this, assume that there is one path $r_j$ that does not pass through any bottleneck. Consider link $i\in E(r_j)$ on the path $r_j$. 
The path flow $f_j$ can be increased by $\delta=\min_{i\in E(r_j)}\{c_i-\sum_{k=1}^\rho B_{i,k}f_{k}\}>0$, such that the new vector of path flows is $f^\prime = (f_1,\ldots,f_j+\delta,\ldots,f_\rho)$. Hence, $f^\prime$ is not proportionally fair because $\sum_{q=1}^{\rho}(f^\prime_q-f_q)/f_q=\delta/f_j>0$, and the path flow $f_j$ can be increased.
$\Box$

\subsection{A centralized algorithm for Proportional Fairness}

%%%% FIGURE %%%%%%%
%%%%%% /FIGURE/ %%%%%%%

%%%% FIGURE %%%%%%%
%%%%%% /FIGURE/ %%%%%%%

Now in order to find the proportionally fair allocation, we need to maximize $U(\vect f )$, constrained to the vector of path flows being feasible, that is:
%%%% EQUATION %%%%%%%
\begin{equation}
\begin{aligned}
& \underset{\vect f}{\text{maximize}}
&  U(\vect f )&=\sum_{j=1}^\rho \log(f_j) \\
& \text{subject to} \quad
& Bf&\leq c \\
& & f_j&\ge 0,
\label{eq:primal}
\end{aligned}
\end{equation}
%%%% EQUATION %%%%%%%
where the link-path incidence matrix is defined by $B_{ij}=1$ if the link $i$ belongs to the path $r_j$ and $B_{ij}=0$ otherwise, and $c=(c_1,\ldots ,c_\eta)$ is the vector of link capacities. The aggregate utility $U(\vect f )$ is concave and the inequality constraints are convex, and hence the optimization problem~(\ref{eq:primal}) is convex. Thus, any locally optimal point is also a global optimum and we can use results from the theory of convex optimization to solve problem~(\ref{eq:primal}) (see~\cite{SCourant89} and ~\cite{SBall_PCM08} for a brief introduction to Lagrange multipliers, and~\cite{SBoyd04} on convex optimization).
The Lagrangian associated with the optimization problem~(\ref{eq:primal})  is~\cite{SKelly98,STan99}:
%%%% EQUATION %%%%%%%
\begin{equation}
L(\vect f, \vect \mu)=\sum_{j=1}^\rho \log(f_j)+\vect \mu^T(\vect c-B\vect f)
\label{eq:Lagrangian}
\end{equation}
%%%% EQUATION %%%%%%%
where $\mu=(\mu_1,\ldots,\mu_\eta)$ is a vector of Lagrange multipliers. The Lagrange dual function~\cite{SBoyd04} is then given by $\sup_{f}L(f,\mu)$, which is easily determined analytically by $\partial L(f^\ast,\mu^\ast)/\partial f=0$ as
%%%% EQUATION %%%%%%%
\begin{align}
\frac{\partial L(f^\ast,\mu^\ast)}{\partial f_{j}^\ast} & =\frac{1}{f_{j}^\ast}-\sum_{i=1}%
^\eta B_{ij}\mu_i^\ast=0\Leftrightarrow \nonumber  \\ 
f_{j}^\ast & =\frac{1}{\sum_{i=1}^\eta B_{ij}\mu_i^\ast},
\label{eq:Lagrange_dual}
\end{align}
%%%% EQUATION %%%%%%%
and thus
%%%% EQUATION %%%%%%%
\begin{equation}
\sup_{f}L(f,\mu)=-\sum\limits_{j=1}^\rho\log\left(  \sum\limits_{i=1}^\eta B_{ij}\mu_i\right)
+\sum\limits_{i=1}^\eta \mu_ic_{i}-\rho
\label{eq:Lagrange_dual_final}
\end{equation}
%%%% EQUATION %%%%%%%

After removing the constant term in equation~(\ref{eq:Lagrange_dual_final}) and converting to a maximization problem, we obtain the dual problem~\cite{SKelly98,STan99}
%%%% EQUATION %%%%%%%
\begin{equation}
\begin{aligned}
& \underset{\vect L}{\text{maximize}}
& V(\mu) &= \sum\limits_{j=1}^\rho\log\left(  \sum\limits_{i=1}^\eta B_{ij}\mu_i\right)
-\sum\limits_{i=1}^\eta \mu_ic_i \\
& \text{subject to} \quad
& \mu_i&\geq 0.
\label{eq:dual}
\end{aligned}
\end{equation}
%%%% EQUATION %%%%%%%

%%%% FIGURE %%%%%%%
\begin{figure*}
\centering
\subfigure{\includegraphics[width=0.8\linewidth]{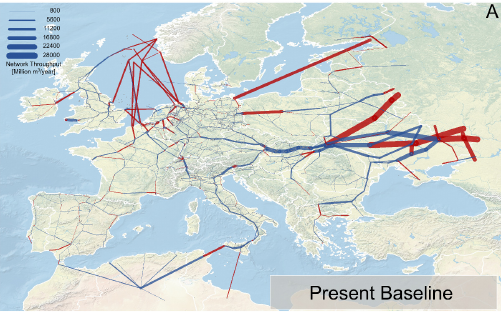} }\\
\subfigure{\includegraphics[width=0.44\linewidth]{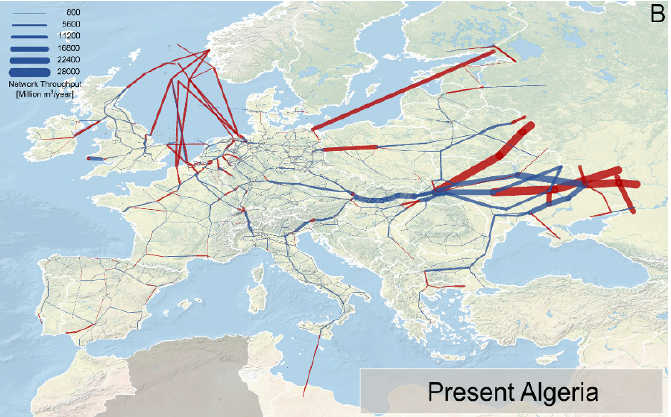} }\quad
\subfigure{\includegraphics[width=0.44\linewidth]{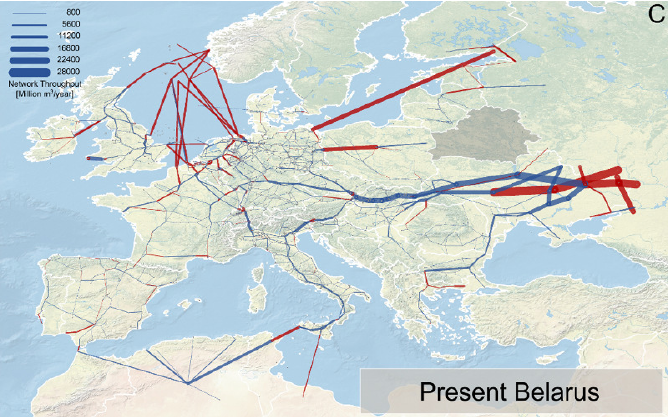} }
\subfigure{\includegraphics[width=0.44\linewidth]{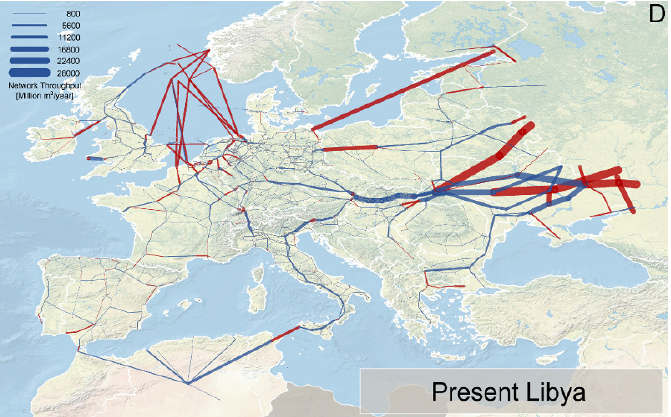} }\quad
\subfigure{\includegraphics[width=0.44\linewidth]{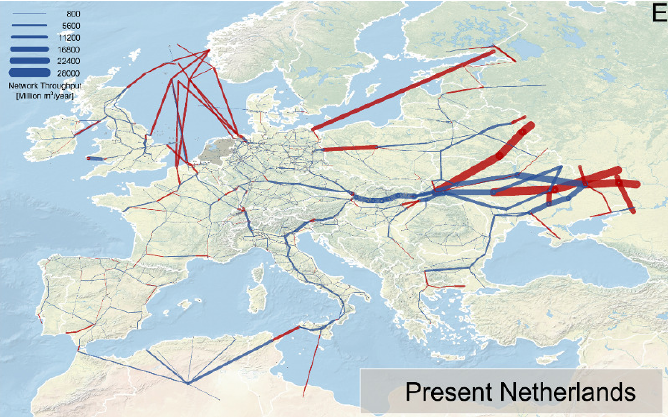}}
\caption{\label{fig:present_scenarios_1} Map of the flow allocation in the following scenarios: (A) present network, (B) present network without Algeria, (C) present network without Belarus, (D) present network without Libya, and (E) present network without the Netherlands. Link thickness is proportional to the total flow on the link. Links in dark red are bottlenecks and links in blue are not used to their full capacity. The scenario country, which is hypothetically removed from the network, is highlighted in gray on the map.}
\end{figure*}
%%%%%% /FIGURE/ %%%%%%%

The primal problem~(\ref{eq:primal}) is convex and the inequality constraints are affine. Hence, Slater's condition is verified and thus strong duality holds. This means that the \textit{duality gap}, \ie~the difference between the optimal of the primal problem~(\ref{eq:primal}) and the optimal of the dual problem~(\ref{eq:dual}), is zero~\cite{SBoyd04}. Strong duality has potentially immense implications as, depending on the problem, it may be easier to solve the primal or the dual. In our case, the primal objective function depends on $\rho$ variables (the path flows) and is constrained by an affine system of equations, whereas the dual objective function depends on $\eta$ variables (the links) and is constrained only by the condition that the Lagrange multipliers are non-negative. Taken together, the methods of convex optimization provide us with powerful tools to gain insights into patterns of congestion in networks where the number $\rho$ of transport routes can be considerably larger than the number $\eta$ of available transport links. Strong duality then states that the optimal path flows $f^\ast$ are related to the optimal Lagrange multipliers $\mu^\ast$ by equation~(\ref{eq:Lagrange_dual}).

Since $f^\ast$ maximizes the Lagrangian over $f$, it follows that its gradient must vanish at $f^\ast$, and thus the following \textit{KKT} condition is satisfied: 
%%%% EQUATION %%%%%%%
\begin{equation}
\mu_{i}^\ast\left(  c_{i}-\sum_{j=1}^\rho B_{ij}f_{j}^\ast\right) =0. 
\label{eq:KKT_complementary_slackness}
\end{equation}
%%%% EQUATION %%%%%%% 
Equation~(\ref{eq:KKT_complementary_slackness}), often referred to as \textit{complementary slackness}~\cite{SBoyd04}, states that the vector $\mu$ of the Lagrange multipliers and the vector of residual capacity have complementary sparsity patterns. To be more specific, either link $i$ is utilized to full capacity (\ie~$c_{i}=\sum_{j=1}^\rho B_{ij}f_{j}^\ast$) and $\mu_i^\ast>0$, or $\mu_i^\ast=0$ and the capacity of link $i$ is underused (\ie~$\sum_{j=1}^\rho B_{ij}f_{j}^\ast<c_{i}$). This gives us a simple and powerful way to identify bottleneck links numerically, as the links with a positive Lagrange multiplier $\mu_i^\ast$.

%%%% FIGURE %%%%%%%
\begin{figure*}
\centering
\subfigure{\includegraphics[width=0.44\linewidth]{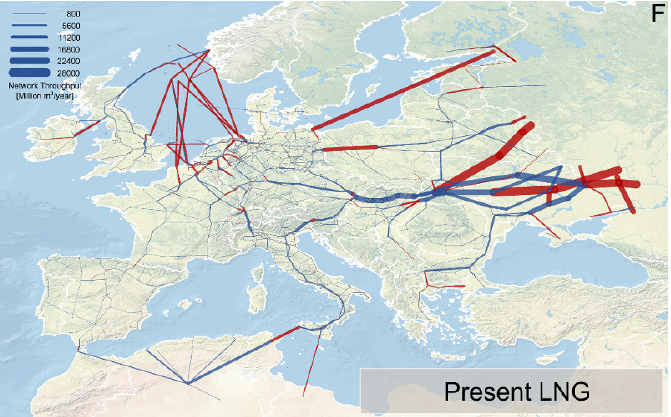} }
\subfigure{\includegraphics[width=0.44\linewidth]{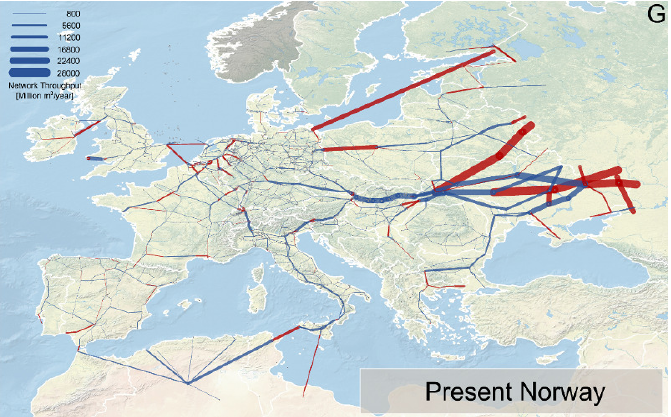} }\quad
\subfigure{\includegraphics[width=0.44\linewidth]{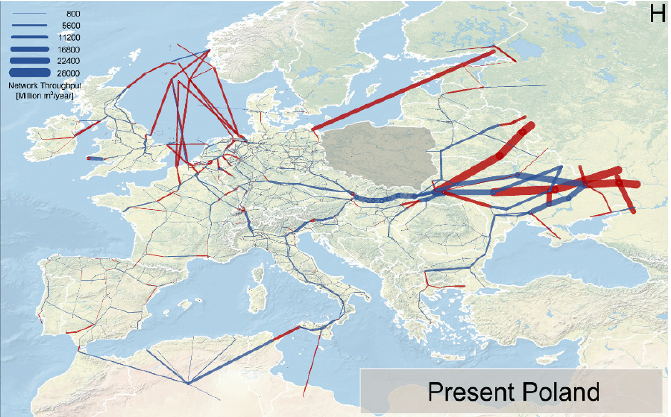} }
\subfigure{\includegraphics[width=0.44\linewidth]{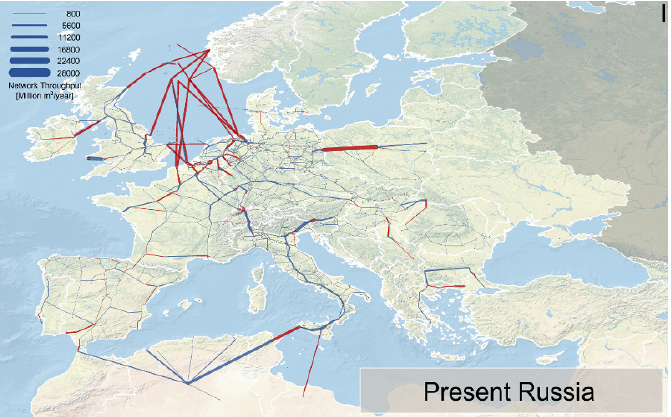} }\quad
\subfigure{\includegraphics[width=0.44\linewidth]{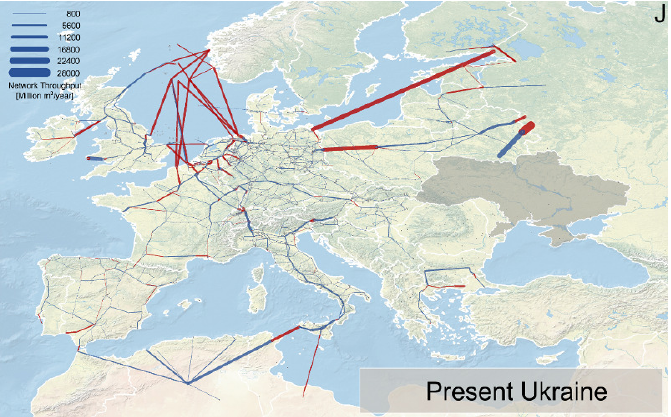} }
\caption{\label{fig:present_scenarios_2} Map of the flow allocation in the following scenarios: (F) present network without Liquefied Natural Gas (LNG) terminals, (G) present network without Norway, (H) present network without Poland, (I) present network without Russia, (J) present network without Ukraine. Link thickness is proportional to the total flow on the link. Links in dark red are bottlenecks and links in blue are not used to their full capacity. The scenario country, which is hypothetically removed from the network, is highlighted in gray on the map.}
\end{figure*}
%%%%%% /FIGURE/ %%%%%%%

\subsection{A centralized algorithm for Proportional Fairness with link price}
Now suppose that the network operator charges a price per unit flow $p_i(y)$  for the use of link $i$, when the total load on the link is $y=\sum_{j=1}^\rho B_{ij}f_j$. This means that the price at one link depends on all the paths that pass through the link~\cite{STan99}. Hence, the problem~(\ref{eq:primal}) can be generalized by adding a cost or penalty that is a function of the price~\cite{SKelly98,STan99}. If the penalty is infinite when the link capacity is exceeded, $y>c_i$, then we can generalize problem~(\ref{eq:primal}) to replace the capacity constraints by the link cost, such that
%%%% EQUATION %%%%%%%
\begin{equation}
\begin{aligned}
& \underset{\vect f}{\text{maximize}}
& \widehat{U}(f,p,y)&=\sum_{j=1}^\rho \log(f_j) -\sum_{i=1}^\eta \int_0^{y_i}p_i(z)dz\\
& \text{subject to} \quad
&  Bf&= y\\
& &  f_j,y_i&\ge 0. 
\label{eq:primal_star}
\end{aligned}
\end{equation}
%%%% EQUATION %%%%%%%

%%%% FIGURE %%%%%%%
\begin{figure*}
\centering
\subfigure{\includegraphics[width=0.8\linewidth]{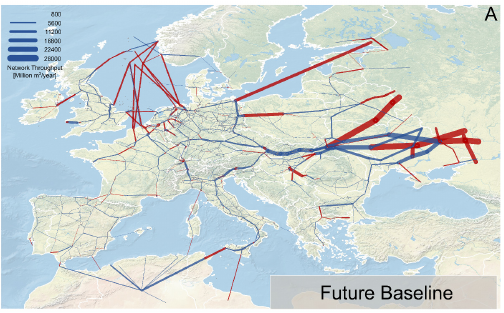} }
\subfigure{\includegraphics[width=0.44\linewidth]{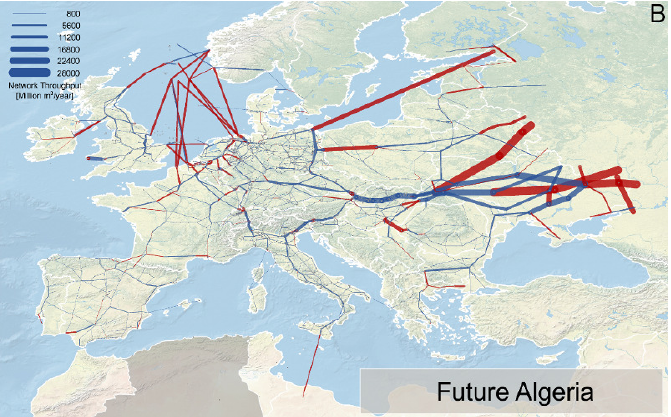} }\quad
\subfigure{\includegraphics[width=0.44\linewidth]{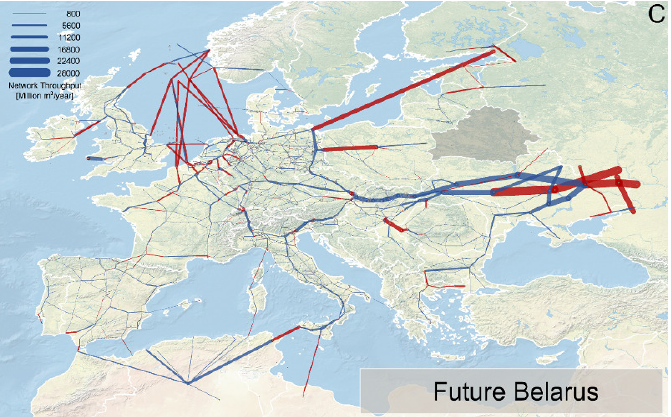} }
\subfigure{\includegraphics[width=0.44\linewidth]{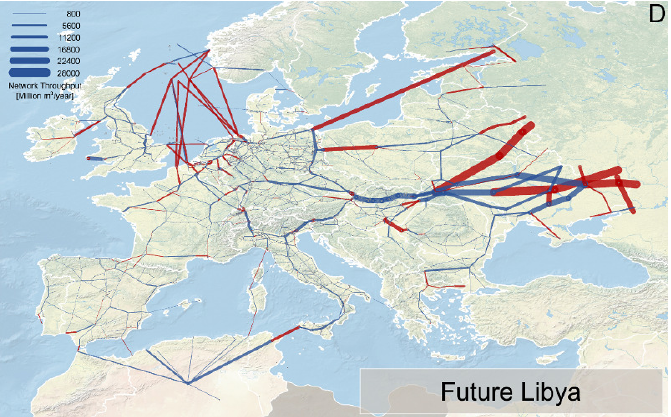} }\quad
\subfigure{\includegraphics[width=0.44\linewidth]{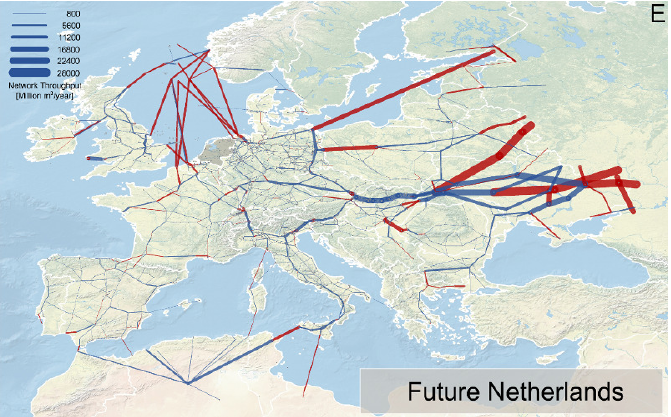}}
\caption{\label{fig:future_scenarios_1} Map of the flow allocation in the following scenarios: (A) future network, (B) future network without Algeria, (C) future network without Belarus, (D) future network without Libya, and (E) future network without the Netherlands. Link thickness is proportional to the total flow on the link. Links in dark red are bottlenecks and links in blue are not used to their full capacity. The scenario country, which is hypothetically removed from the network, is highlighted in gray on the map.}
\end{figure*}
%%%%%% /FIGURE/ %%%%%%%

To derive the dual of problem~(\ref{eq:primal_star}), we first find its Lagrange dual
%%%% EQUATION %%%%%%%
\begin{equation}
\begin{aligned}
\widehat{g}(\mu) =& \sum_{j=1}^\rho \log\left(\frac{1}{\sum_{i=1}^\eta B_{ij}\mu _i}\right)-\rho+\\
&\sum_{i=1}^\eta\left ( \mu_i p_i^{-1}(\mu_i)-\int_0^{p_{i}^{-1}\left( \mu_i\right) }p_i(z) dz\right )
\label{eq:network*_Lagrange_dual_long}
\end{aligned}
\end{equation}
%%%% EQUATION %%%%%%%
To simplify equation~(\ref{eq:network*_Lagrange_dual_long}), we integrate by parts and then by substitution, 
%%%% EQUATION %%%%%%%
\begin{equation}
\widehat{g}(\mu) = \sum_{j=1}^\rho \log\left(\frac{1}{\sum_{i=1}^\eta B_{ij}\mu_i}\right)-\rho+\sum_{i=1}^\eta \int_0^{\mu_i}q_i(x)dx,
\label{eq:network*_Lagrange_dual}
\end{equation}
%%%% EQUATION %%%%%%%
where $q(\cdot )$ is the inverse of $p(\cdot )$. Following ~\cite{SKelly98,STan99}, we now remove the constant term in equation~(\ref{eq:network*_Lagrange_dual}) and covert to a maximization problem to obtain the dual of problem~(\ref{eq:primal_star}):
%%%% EQUATION %%%%%%%
\begin{equation}
\begin{aligned}
& \underset{\vect L}{\text{maximize}}
& \widehat{V}(\mu, q) &= \sum_{j=1}^\rho \log\left(\sum_{i=1}^\eta B_{ij}\mu_i\right)-\sum_{i=1}^\eta \int_0^{\mu_i}q_i(x)dx\\
& \text{subject to} 
& & \mu_i\ge 0. 
\label{eq:dual*}
\end{aligned}
\end{equation}
%%%% EQUATION %%%%%%%
%%%% FIGURE %%%%%%%
\begin{figure*}
\centering
\subfigure{\includegraphics[width=0.44\linewidth]{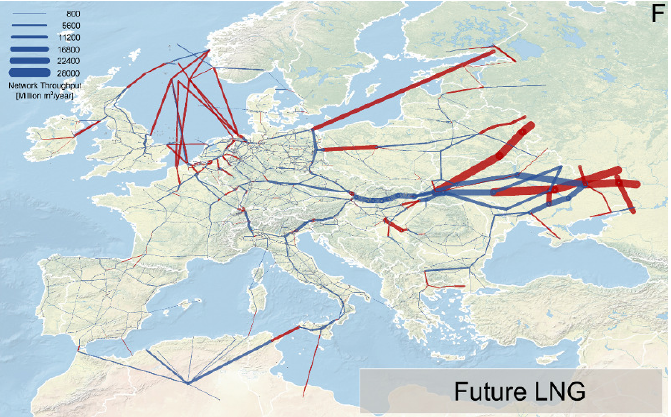} }\quad
\subfigure{\includegraphics[width=0.44\linewidth]{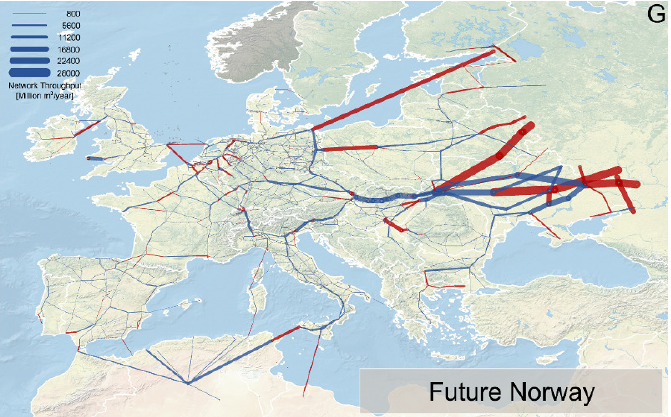} }
\subfigure{\includegraphics[width=0.44\linewidth]{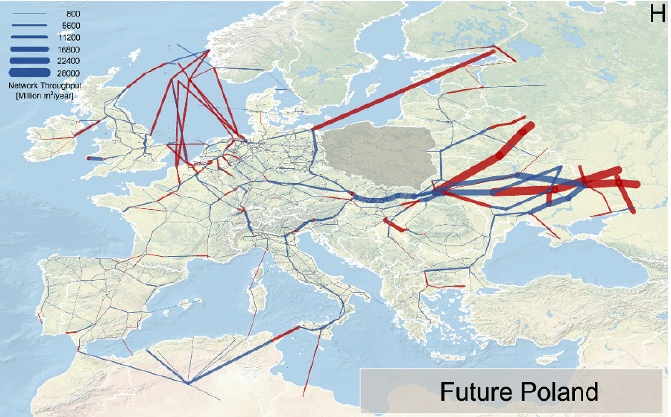} }\quad
\subfigure{\includegraphics[width=0.44\linewidth]{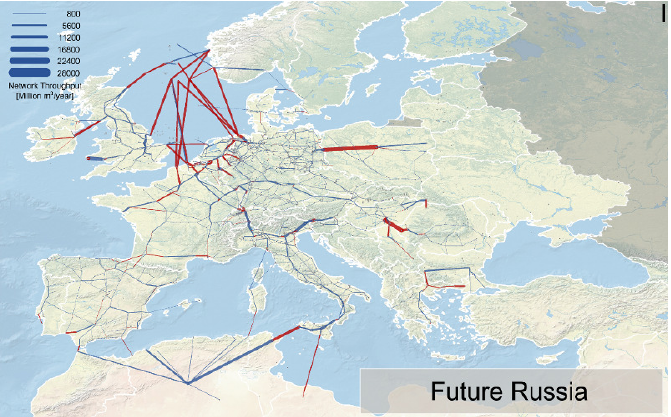} }
\subfigure{\includegraphics[width=0.44\linewidth]{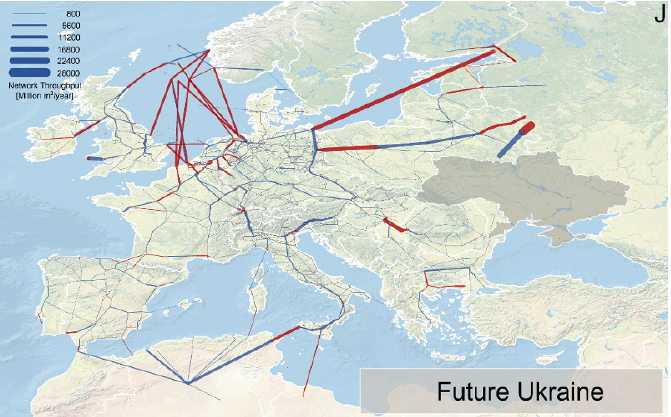} }
\caption{\label{fig:future_scenarios_2} Map of the flow allocation in the following scenarios: (F) future network without Liquefied Natural Gas (LNG) terminals, (G) future network without Norway, (H) future network without Poland, (I) future network without Russia, (J) future network without Ukraine. Link thickness is proportional to the total flow on the link. Links in dark red are bottlenecks and links in blue are not used to their full capacity. The scenario country, which is hypothetically removed from the network, is highlighted in gray on the map.}
\end{figure*}
%%%%%% /FIGURE/ %%%%%%%

The dual problem~(\ref{eq:dual*}) is equivalent to the original dual problem~(\ref{eq:dual}) if $q_i(x)=c_i$. However, this function is non-invertible and thus we approximate it by the invertible function~\cite{SKelly98,STan99}
%%%% EQUATION %%%%%%%
\begin{equation}
q_i(x)= \frac{x c_i}{x + \epsilon}.
\label{eq:q_OR}
\end{equation}
%%%% EQUATION %%%%%%%
Problems~(\ref{eq:dual*}) and (\ref{eq:dual}) are thus equivalent in the limit $\epsilon\rightarrow 0$. The primal problems~(\ref{eq:primal}) and (\ref{eq:primal_star}) are equivalent in the limit $\epsilon\rightarrow 0$ if
%%%% EQUATION %%%%%%%
\begin{equation}
p_i(y)\sim q_i^{-1}(y)=y \epsilon/(c_i-y).
\label{eq:p_OR}
\end{equation}
%%%% EQUATION %%%%%%%

\subsection{A decentralized algorithm for Proportional Fairness}
The key to a decentralized algorithm for Proportional Fairness is to translate the problem~(\ref{eq:primal_star})  into an autonomous system of coupled differential equation, with a fixed point equivalent to the optimal solution of the optimization problem. To do this, we use the result that the stable fixed point of a system of differential equations is the maximum of the equations' Lyapunov function

For each path $r_j$, the network is offering a certain path flow $f_j$ with unit rate of change, $df_j/dt=1$. Now suppose that the network operator charges a price per unit flow $p_i(y)$  for the use of link $i$, when the total load on the link is $y=\sum_{j=1}^\rho B_{ij}f_j$. This means that the price at one link depends on all the paths that pass through the link. The price causes a reduction in the path flow $f_j$, such that 
%%%% EQUATION %%%%%%%
\begin{equation}
\frac{d}{dt}f_j(t) =1-f_j(t) \sum_{i=1}^\eta B_{ij}\mu_i(t),
\label{eq:diff_eqs_primal}
\end{equation}
%%%% EQUATION %%%%%%% 
where the price on link $i$ is 
%%%% EQUATION %%%%%%%
\begin{equation}
\mu_i(t)=p_i\left (\sum_{j=1}^\rho B_{ij}f_j(t)\right ).
\label{eq:mu_total_price}
\end{equation}
%%%% EQUATION %%%%%%%
User $i$ responds to an underused capacity with a steady increase of its path flow, and to congestion with a multiplicative decrease of its path flow at a rate proportional to the congestion price. This additive-increase/multiplicative-decrease mechanism is best known for its use in communication networks, and is implemented in TCP congestion avoidance~\cite{SJacobson88,SChiu_and_Jain89}.

A possible pricing policy consists in charging only for link flows that are close to capacity, with a sharp increase in the price that each path pays as the link becomes saturated:
%%%% EQUATION %%%%%%%
\begin{equation}
p_i(y) = \frac{\max (0,y-c_i+\epsilon)}{\epsilon ^2}.
\label{eq:price}
\end{equation}
%%%% EQUATION %%%%%%%
As $\epsilon\rightarrow 0$, the price $p_i$ tends to zero for link flows below capacity, and to infinity for saturated links. Hence, problem~(\ref{eq:primal_star}) approximates arbitrarily closely the primal problem~(\ref{eq:primal}).

%%%% TABLE %%%%%%%
\begin{table*}[htbp]
\centering
\label{table:from_data_network}
{\bf Table 1. Summary of notation\\}
\begin{tabular}{ll}
\hline
{\small Indexes:} &\\ 
$i$ & Link index\\
$j$ & Path index\\
$m$ & Exporting country\\
$n$ & Importing country\\
$k$ & Node index in an exporting country\\
$l$ & Node index in an importing country\\
\hline
{\small Extracted at network level:} & \\

$\eta$ & Number of links in the network\\
$d_i$ & Diameter of link $i$ (units: $\mathrm{in}$;  vector with dimension $\eta$)\\
$c_i=0.56d_i^{2.5}$ & Capacity of link $i$. The exponent $2.5$ is found from data~\cite{SCarvalho09}, and the prefactor\\
& is obtained by calibrating the present baseline scenario to match the $T_{mn}$\\
& flow matrix (units: $\mathrm{Mm^3/year}$; vector with dimension $\eta$)\\
$l_i$ & Length of link $i$  (units: $\si{\km}$; vector with dimension $\eta$)\\

$B_{ij}$ & Link-path incidence matrix. $B_{ij}=1$ if the link $i$ belongs to the path $r_j$ and\\
&$B_{ij}=0$ otherwise (matrix with dimensions $\eta \times \rho$)\\
$b_i=\sum_{j=1}^\rho B_{ij}$ & Number of paths that pass through link $i$ (vector with dimension $\eta$)\\

\hline
{\small Extracted at country level:} & \\
$\nu$ & Number of countries\\
$s_m$ & Number of (source) nodes in an exporting country $m$\\
$t_n$ & Number of sink nodes in an importing country $n$\\
$g_n$ & Number of Liquefied Natural Gas (LNG) terminals in an importing country $n$\\

$T_{mn}$ & Volume of gas received by an importing country $n$ from an exporting country $m$\\
&and LNG (matrix with dimensions: $(\eta+1)\times \eta$)\\
$\widehat{T}_{mn}=\left\{\begin{array}
{ll}
1 & \text{if }T_{mn}\text{ is positive}\\
0 & \text{otherwise}
\end{array}
\right.$ & Binary matrix with zero entries if there is no transport between countries $m$ and $n$\\
$Z_{nl}$ & Population associated with sink node $l$ of importing country $n$\\
&(matrix with dimensions $\nu\times$ [Number of Voronoi and urban sinks])\\
$z_n$ & Population of an importing country $n$\\

\hline
{\small Computed for the routing:} &\\ 
$\rho$ & Number of paths on the network\\
$r_{m,k,n,l}$ & Path connecting source node $k$ in an exporting country $m$ with sink node $l$\\
&in an importing country $n$, where $k=1,\dots,\Phi_{mn}$ if $m$ is an exporting country\\
&and $k=1,\dots,g_n$ if $m$ is LNG\\
$r_j$ & Paths are also indexed by an integer, to simplify the notation\\
&used in the congestion control algorithm (vector with dimension $\rho$)\\
$h_i=c_i/(1+b_i)$ & Share of capacity allocated to each path passing through link $i$\\
&at the beginning of the heuristic rerouting\\ 
$D_{mnl}$ & Demand of sink $l$ in an importing country $n$ satisfied by an exporting country $m$\\
$\overline{D}_{mnl}=\left \lfloor \frac{D_{mnl}}{\min(D_{mnl})}\right \rfloor$ & Number of identical paths between a source and sink pair,\\
&each having demand $\min(D_{mnl})$\\
$\widetilde{l_i}=\left(  \frac{\left\langle h_i\right\rangle }{h_i}\right)^{\alpha}l_i$ & Effective length of link $i$\\
$\overleftrightarrow{l_j}$ & Effective length of path $j$ (vector with dimension $\rho$)\\

\hline
{\small Parameters:} &\\ 
$\alpha=0.03$ &Exponent of $\langle h\rangle /h$ in equation~(\ref{eq:link_weights}) (see also Figure~\ref{fig:ALPHA_parameter})\\
$\Phi_{mn}=\left\{\begin{array}
{ll}
\min(10,s_m) & \text{if }m\text{ is a country}\\
g_n & \text{if }m\text{ is LNG }
\end{array}
\right.$ & If $m$ is a country, we connect each sink node to the $\min(10,s_m)$\\
& geographically closest nodes in an exporting country $m$\\
&(distance measured along network paths); if $m$ is LNG, we connect\\
& each sink node to the $g_n$ LNG terminals in an importing country $n$.\\

\hline
{\small Congestion control algorithm:} &\\ 
$f_j$ & Path flow on path $j$ (dimension $\rho$)\\
$\mu_i$ & Price on link $i$, and dual of $f_j$ (vector with dimension $\eta$)\\
$p_i$ & Price function on link $i$ (vector with dimension $\eta$)\\
$q_i$ & Inverse of $p_i$ (vector with dimension $\eta$)\\
$L(f,\mu)$ & Lagrangian function of the primal problem\\
$V(\mu)$ & Lagrangian function of the dual problem\\
 
\end{tabular}
\end{table*}
%%%% TABLE %%%%%%%

\section{Results}
Figures~\ref{fig:present_scenarios_1}, \ref{fig:present_scenarios_2}, \ref{fig:future_scenarios_1} and~\ref{fig:future_scenarios_2} show the load on network links in the present and future scenarios. These figures demonstrate that our model reproduces the main transport corridors in Europe, and show the spatial pattern of bottleneck links for each scenario. It is apparent how a hypothetical removal of either Russia or Ukraine cuts-off the major transport routes, and damages drastically the supply of populations in Europe.

\subsection{Detailed interpretation of results at country and urban levels}
At country level:
\begin{itemize}
\item Greece receives its gas from diverse sources, and thus is resilient to the scenarios we analyse. It gets most of its gas from Russia ($65.2\%$), from Turkey ($16.4\%$)  and LNG ($18.4\%$);
\item Ireland gets its gas from the UK and is unaffected by our scenarios, since the UK is a secure source in our model;
\item Switzerland acts like a hub between South and Northern Europe, so it has very good access to network capacity;
\item Ukraine is a major transit route for gas to Europe;
\item Latvia and Finland have a relatively small population, good access to network capacity and are very close to Russia;
\item Surprisingly, Belarus does better when Ukraine is removed from the network. The apparent contradiction is solved by realizing that Europe's supply from Russia has been historically built around Ukraine. Hence, Ukrainian routes have higher capacity and shorter routes to central Europe than routes that pass through Belarus. In contrast, when Ukraine is removed from the network, routes through Belarus become the first choice to supply central Europe;
\item Belgium  draws its high energy security from diversification of supply. It gets its gas from the Netherlands ($42.5\%$), from Norway($48.7\%$, including LNG), Russia ($2.8\%$), Germany ($1.2\%$), and the UK ($4.8\%$).
\end{itemize}

At urban level:
\begin{itemize}
\item Surprisingly, Rome seems to gain slightly from removing Libya. Rome is approximately in the middle of the Italy, and the country is supplied both from the South and from the North. When Libya is removed there is no need to transport gas from the South to the North of Italy and this frees  capacity to bring more gas from the North to Rome;
\item Unexpectedly, Berlin gains from the removal of Poland. Germany is transporting gas to Poland. When Poland is removed, the capacity that is freed can be used for German cities located close to the Polish border, including Berlin; \item Finally, Dublin is resilient to all scenarios because it gets all of its supply from the UK, and we do not have any scenario affecting the ability of UK to supply gas.
\end{itemize}
%% == end of paper:

\end{document}